\documentclass[11pt,a4paper]{article}
\usepackage{fullpage}
\usepackage{amsmath, amsthm, amsfonts, amssymb, latexsym}
\usepackage{mathtools} 
\let\originalleft\left
\let\originalright\right
\renewcommand{\left}{\mathopen{}\mathclose\bgroup\originalleft}
\renewcommand{\right}{\aftergroup\egroup\originalright}

\usepackage[colorlinks, citecolor=Red, linkcolor=Blue, urlcolor=blue]{hyperref}
\usepackage{caption}
\usepackage[dvipsnames]{xcolor}
\usepackage{tikz}
\usepackage{enumerate}
\usepackage{cancel}
\usepackage{dsfont}
\usepackage{accents}
\usepackage{tensor}

\usepackage{authblk}

\newcommand{\rhs}{r.h.s.\ }
\newcommand{\lhs}{l.h.s.\ }
\newcommand{\wrt}{w.r.t.\ }
\newcommand{\cf}{cf.\ }

\newcommand{\ibar}{{\frac{\mathrm{i}}{\hbar}}}

\makeatletter
\DeclareRobustCommand{\lambdabar}{\mathord{%
 \text{$\m@th\mkern+1mu\raisebox{-0.2ex}[0pt][0pt]{$\mathchar'26$}\mkern-10mu \lambda$}%
}}
\makeatother

\newcommand{\eti}[1]{\mathrm{e}_\otimes^{\ibar #1}}
\newcommand{\et}[1]{\mathrm{e}_\otimes^{#1}}

\newcommand{\ud}{\mathrm{d}}
\newcommand{\del}{\partial}

\newcommand{\R}{\mathbb{R}}

\newcommand{\F}{\mathfrak{F}}

\newcommand{\1}{\mathbbm{1}}

\newcommand{\cR}{\mathcal{R}}

\newcommand{\cD}{\mathcal{D}}

\newcommand{\cT}{\mathcal{T}}

\newcommand{\eps}{\varepsilon}

\def\1{{\mathds{1}}}

\newcommand{\ia}{{\mathrm{int}}}
\newcommand{\ret}{{\mathrm{r}}}
\newcommand{\adv}{{\mathrm{a}}}

\newcommand{\loc}{{\mathrm{loc}}}
\newcommand{\feyn}{{\mathrm{F}}}
\newcommand{\weyl}{{\mathrm{W}}}

\newcommand{\vp}{{\varphi}}

\newcommand{\nn}{\nonumber}
\newcommand{\beq}{\begin{equation}}
\newcommand{\eeq}{\end{equation}}

\newcommand{\defeq}{\mathrel{\coloneq}}

\newcommand{\vol}{\mathrm{vol}}

\newcommand{\os}{{\mathrm{o.s.}}}

\DeclareMathOperator{\supp}{supp}
\DeclareMathOperator{\Deg}{Deg}

\newcommand{\cO}{{\mathcal{O}}}

\newcommand{\cE}{{\mathcal{E}}}

\newcommand{\cevvec}[1]{\accentset{\leftrightarrow}{#1}}

\newcommand{\rmi}{\mathrm{i}}

\begin{document}

\title{The Weyl anomaly in interacting quantum field theory on curved spacetimes}
\author{Markus B.~Fr\"ob\thanks{mfroeb@itp.uni-leipzig.de} \ and Jochen Zahn\thanks{jochen.zahn@itp.uni-leipzig.de} \\ Institut f\"ur Theoretische Physik, Universit\"at Leipzig\\ Br\"uderstr.\ 16, 04103 Leipzig, Germany}

\date{\today}

\maketitle

\begin{abstract}
We define the notion of Weyl anomalies, measuring the violation of local scale invariance, in interacting quantum field theory on curved spacetimes in the framework of locally covariant field theory. We discuss some general properties of Weyl anomalies, such as their relation to the trace anomaly. We give a criterion for a theory to be conformal at the quantum level, and show that even for a conformal theory Weyl transformations in general obtain quantum corrections. We study the trace anomaly in detail for the $\phi^4$ theory, in particular determining it up to second order in the interaction. We also show that at third order in the interaction a potential $\Box \phi^2$ term can be removed by finite renormalization.
\end{abstract}

\section{Introduction}

A classical field theory whose action is invariant under local scale transformations has a traceless stress tensor $T_{\mu \nu}$. However, this is in general no longer the case for the corresponding quantum field. One calls this a \emph{trace anomaly}. The trace anomalies of free theories (in four spacetime dimensions) have been determined about 50 years ago \cite{CapperDuff1974, Christensen76, Brown1977, ChristensenFulling77, DowkerCritchley, Duff1977, WaldTraceAnomaly} (see also \cite{DuffReview1993} for a review and \cite{1702.04247, CasarinGodazgarNicolai, BastianelliChiese, FilocheLarueQuevillonVuong, CorianoLionettiMaglio, LarueQuevillonZwicky, FerreroEtAlTraceAnomaly, 2312.13222, 2406.12464, 2407.00415, 2411.03842} for recent work) and are well-established (even though there were debates concerning chiral fermions until recently \cite{Nakayama2012, Bonora2014, Bastianelli2016, FrobZahn2019, AbdallahEtAl2021}): They are of the form (the symbol $\cT$ stands, in the present context, for a locally covariant ``Wick ordering'' mapping classical local fields to quantum fields, as explained below)
\begin{equation}
\label{eq:FreeTraceAnomalyGeneral}
 \cT( g^{\mu \nu} T_{\mu \nu} ) = - a \cE_4 + c C^2 + b \Box R \,,
\end{equation}
with $\cE_4$ the Euler density, $C^2$ the square of the Weyl tensor, and $a$, $b$, $c$ real coefficients of $\cO(\hbar)$. While $a$ and $c$ are, for a given theory, fixed, $b$ is subject to renormalization ambiguities, which can be used to set $b=0$.\footnote{In principle, such a renormalization freedom must be fixed by experiments. However, there are strong arguments that a nonzero $b$ is physically unacceptable \cite{HorowitzWald1978}.} That these are generic features of free theories follows from cohomological arguments \cite{BonoraEtAl83}.

Soon after the trace anomaly in free theories was worked out, attention began to focus on interacting theories. The $\phi^4$ model \cite{BrownCollins, HathrellScalar}, QED \cite{HathrellQED}, non-abelian gauge theories \cite{Freeman1983, Osborn1989}, and general renormalizable models combining these \cite{JackOsborn}, were considered using dimensional regularization and renormalization group methods. 
These allowed to reduce the computations (up to some order in the loop expansion) to calculations on flat space. 
Specifically, for the $\phi^4$ model it was found \cite{BrownCollins, HathrellScalar} that the trace anomaly acquires, apart from the conformally invariant terms $\phi^4$ and $\phi ( \Box - \frac{1}{6} R ) \phi$, also non-invariant terms $\Box \phi^2$, $R \phi^2$, $R^2$, and $\Box R$. 
Specifically, the term $\Box \phi^2$ is found to be of $\cO(\lambda^3)$, the terms $R \phi^2$ and $\Box R$ of $\cO(\lambda^4)$, and the term $R^2$ of $\cO(\lambda^5)$, with $\lambda$ the coupling constant. Furthermore, the anomaly coefficients $a$, $c$ of the free theory receive corrections of $\cO(\lambda^4)$ and $\cO(\lambda^2)$, respectively.

A complementary approach to trace anomalies in interacting theories was initiated in \cite{BonoraEtAl83}, where the general form of the trace anomaly was investigated cohomologically, independently of regularization schemes, in a path integral framework. It was argued that the conformally non-invariant terms $\Box \phi^2$, $R \phi^2$, $R^2$, and $\Box R$ can either not occur or can be removed. The contradiction with the results of \cite{BrownCollins, HathrellScalar} mentioned above is resolved by recognising that the arguments of \cite{BonoraEtAl83} in fact only apply to the leading order in $\hbar$. To go beyond leading order, variable coupling ``constants'' have to be taken into account, as shown in \cite{Osborn1991}. There, all possible terms in the trace anomaly containing derivatives of the coupling constants are taken into account, and consistency conditions are used to derive relations between these. 
Also the renormalization ambiguity of the trace anomaly is characterized. 
The results both of the concrete calculations, in particular \cite{Osborn1989, JackOsborn}, as well as the general framework developed in \cite{Osborn1991}, played an important role in validations and discussions of the $a$-theorem \cite{Cardy1988, KomargodskiSchwimmer, Fortin:2012hn}.

The results on the trace anomaly in interacting theories discussed above are all based on the generating functional $W$ of connected correlation functions. From the point of view of quantum field theory on curved spacetimes, this starting point has conceptual disadvantages: While in a Riemannian context (which is implicitly used in the above mentioned works) there is a preferred choice of correlator, this is no longer true for general spacetimes. For example, on the exterior region of Schwarzschild spacetime, the Boulware, Unruh, and Hartle-Hawking state are all well-defined and sensible, but describe very different physical situations. One may like to avoid expressing an important structural property like the breaking of conformal invariance in terms of a contingent quantity like $W$.

In recent decades, locally covariant field theory \cite{HollandsWaldWick, HollandsWaldTO, BrunettiFredenhagenVerch, HollandsWaldReview} has been established as a mathematically rigorous and conceptually clear framework for the description of perturbative quantum field theory on curved spacetimes. It is based on the algebraic approach \cite{HaagBook}, so it does not rely on specific states or Hilbert space representations and ensures the correct transformation of local observables (``renormalized operators'') under isometries. It is independent of particular renormalization schemes, but has an intrinsic notion of renormalization group flow \cite{HollandsWaldRG}.

Here, we perform some first steps for the description and study of (the breaking of) conformal invariance in the framework of locally covariant field theory.\footnote{Recently, a different approach to the trace anomaly in interacting theories was pursued in \cite{CosteriDappiaggiGoi}. We comment on the relation to our approach below.} The starting point is a free action $S_0$ exhibiting local scale invariance.\footnote{This excludes gauge theories, whose local scale invariance is broken by gauge fixing. However, as the breaking terms are BRST exact (and in this sense controllable), we expect that our formalism can be extended to encompass also gauge theories.} 
The crucial ingredient for the definition of interacting fields are then time-ordered products $\cT(F_1 \otimes F_2 \otimes \dots)$, which take local functionals $F_i$ as arguments.\footnote{In order to formulate the axioms for Wick powers and time-ordered products in a coherent manner, it is convenient to express Wick powers as time-ordered products with a single factor.} In order to properly define such time-ordered products, certain distributions of several spacetime variables, which are defined only up to coinciding points, i.e., on $M^k \setminus \{ (x, \dots, x) \mid x \in M \}$ (with $M$ being the spacetime manifold), have to be extended, i.e., defined on $M^k$.\footnote{This is the analog of renormalization in other approaches to quantum field theory.} In this extension process, local conformal invariance is in general broken. 
In analogy to the treatment of gauge symmetries, where the breaking of local gauge invariance is described by an anomaly functional in an anomalous master Ward identity \cite{HollandsYM}, the breaking of conformal invariance can be subsumed in the \emph{Weyl anomaly} $A(F_1 \otimes F_2 \otimes \dots )$, which is itself a local functional. 
The Weyl anomaly is subject to \emph{consistency conditions} which are useful to analyse the possible anomalies.

After defining the Weyl anomaly, we investigate some of its properties. 
We show that the trace anomaly can be understood as a special case of the Weyl anomaly, in the sense that the interacting contribution to the trace anomaly is $A( \et{S_\ia} )$, with $S_\ia$ the interacting part of the action. We also relate the Weyl anomaly to the renormalization group flow, in particular showing that $A( \et{S_\ia} )$ is invariant under this flow. We also investigate the breaking of conformal invariance in interacting theories and give a (tentative) characterization of conformal field theories in our perturbative framework: 
Namely, a classically conformally invariant theory is conformal at the quantum level if the anomaly $A(\et{S_\ia})$ is the sum of a c-number and a term proportional to the classical trace of the stress tensor, which means that the trace anomaly after ``field strength renormalization'' is a c-number. 
In this case, we give an expression for the \emph{quantum Weyl transformation} of observables, which differ from the classical Weyl transformation by corrections of order $\hbar$.

The abstract concept is then applied to the concrete case of the $\phi^4$ model. We explicitly compute $A(\et{S_\ia})$ up to second order in the interaction. We find that:

\begin{itemize}

\item At first order in the interaction, the anomaly $A(S_\ia)$ in a Hadamard point-splitting scheme is proportional to $\Box \phi^2$, which is cohomologically exact and can be removed by a finite renormalization. In fact, one can even fulfill the stronger condition $A( \phi^4 ) = 0$.

\item When the removal of $A(\phi^4)$ is performed, then at second order in the interaction, the trace anomaly is a linear combination of $\cE_4$, $C^2$, $\phi ( \Box - \frac{1}{6} R ) \phi$, $\phi^4$, $\Box \phi^2$, and $\Box R$, the last two of which can be removed by a finite renormalization. The coefficients of $\phi ( \Box - \frac{1}{6} R ) \phi$, $\phi^4$, and $\cE_4$ are fixed (the latter vanishing), while that of $C^2$ (which is of $\cO(\hbar^3)$) is scheme dependent.

\item Had we not removed the anomaly at first order, then the second order anomaly would have also contained the terms $R \phi^2$ and $R^2$.

\item While not explicitly computing the trace anomaly at third order in the interaction, we show that it can only contain the same terms as the second order anomaly, and again $\Box \phi^2$ and $\Box R$ can be removed. Again, it is crucial that we removed the total derivative terms $\Box \phi^2$ and $\Box R$ from the second order anomaly.

\end{itemize}

Our results agree with those obtained in \cite{BrownCollins}, \cite{HathrellScalar} (which extend to higher orders in the interaction), with two exceptions: i) The coefficient of the $\Box \phi^2$ term is found in \cite{HathrellScalar} (and similarly in \cite{BrownCollins}) to be of the form $\eta - d$, with $d = d_3 \lambda^3 + \cO(\lambda^4)$ and $\eta$ subject to an inhomogeneous differential equation with general solution $\eta = ( \eta_3 \lambda^3 + k \lambda^{\frac{1}{3}} ) ( 1 + \cO(\lambda))$, where $\lambda$ is the coupling constant and $k$ a free parameter. Obviously, a non-zero $k$ amounts to incorporating non-perturbative effects, and also seems at odds with the framework of \cite{Osborn1991}, which assumes a smooth dependence on the coupling ``constants''. However, as $d_3 \neq \eta_3$, such non-perturbative effects need to be invoked in order to achieve a vanishing coefficient of $\Box \phi^2$ at third order in the interaction. In our framework, which is fully perturbative, this is not necessary.\footnote{A $\Box \phi^2$ term at $\cO(\lambda^3)$ (without a possibility to remove it), was recently also found in \cite{IrgesKarageorgos}.} ii) Within the dimensional regularization scheme used in \cite{HathrellScalar}, the second order contribution to $C^2$ is not subject to ambiguities. We find the same value in a renormalization scheme which is in some sense ``minimal'', but there seems to be no fundamental reason to use this particular scheme. In our framework, the second order contribution to $C^2$ is ambiguous (in accordance with \cite{Osborn1991}) and must in principle be fixed by experiment.

A further notable difference of our approach and that of \cite{HathrellScalar}, \cite{Osborn1991} is the treatment of terms in the trace anomaly which classically vanish on-shell, namely $\phi ( \Box - \frac{1}{6} R ) \phi - \frac{\lambda}{6} \phi^4$ in $\phi^4$ theory. While such terms seem to be neglected in \cite{Osborn1991} (but see \cite{Fortin:2012hn} for a discussion on how to include these), they are included in \cite{HathrellScalar} but such that their contribution to the on-shell trace anomaly vanishes. 
In contrast, in our approach, a contribution $\gamma[ \phi ( \Box - \frac{1}{6} R ) \phi - \frac{\lambda}{6} \phi^4]$ to the interacting part of the trace anomaly cannot be assumed to vanish in the quantum theory.
However, noting that it is nothing but the classical expression for the trace of the stress tensor, we can interpret this term as implementing a ``field strength renormalization'' of the trace anomaly. It is in this manner that we recover the results of \cite{HathrellScalar} regarding the coefficients of $\cE_4$ and $C^2$ at second order in the interaction.

It may be noteworthy that the determination of the coefficients of $\cE_4$ and $C^2$ at second order in the interaction corresponds to a three-loop calculation. We are not aware of previous calculations to that order on general curved spacetimes.\footnote{In \cite{HathrellScalar} some coefficients of the trace anomaly are computed to even higher order. But, as mentioned above, renormalization group methods are used to deduce these from flat space results.}

An obvious question is whether the above results generalize to yet higher orders in the interaction, i.e., whether one can achieve it to only contain the terms already present at second order. The answer given to this question in the literature is ``no'' \cite{HathrellScalar}. While we do not attempt to verify this in our framework, we relate this question to the presence or absence of certain terms involving derivatives of the coupling ``constant'', in agreement with the results of \cite{Osborn1991}. 

The article is structured as follows: In the next section, we review the framework of locally covariant field theory \cite{HollandsWaldWick, HollandsWaldTO, HollandsWaldStress, HollandsWaldReview}, with one modification: By assuming that the free action is invariant under local scale transformations, we formulate the scaling transformation, which in \cite{HollandsWaldWick, HollandsWaldTO, HollandsWaldStress, HollandsWaldReview} is only defined for a constant scaling factor, for general (non-constant) scaling factors. In Section~\ref{sec:WeylAnomaly} we define the Weyl anomaly as capturing the violation of local scale invariance by the time-ordered products. We also discuss general properties of the Weyl anomaly, relate it to the trace anomaly, and define quantum Weyl transformations. In Section~\ref{eq:phi4_Cohomology}, we study the trace anomaly in $\phi^4$ theory from a cohomological point of view. 
In particular, we show how a vanishing anomaly $A(\phi^4) = 0$ implies that at third order in the interaction terms like $\Box \phi^2$ and $\Box R$ can be removed (and $R \phi^2$ and $R^2$ cannot occur). 
Also the importance of controlling derivatives of the coupling ``constant'' is discussed. In Section~\ref{sec:phi4_Computation}, we explicitly compute the trace anomaly to second order in the interaction. We conclude with a brief summary and an outlook. An appendix contains proofs of some statements mentioned in the main text.

\subsubsection*{Notation and conventions:} 
We are using the ``mostly plus'' convention for the metric and the conventions of \cite{WaldGR} regarding the curvature tensors. 
The d'Alembertian is denoted by $\Box = \nabla^\mu \nabla_\mu$, and $\vol(x) = \sqrt{ - \left| g \right| } \, \ud^4 x$ denotes the volume element. $J^\pm(C)$ denotes the causal future/past of a subset $C \subset M$ of spacetime.

\section{A review of locally covariant field theory}

We want to establish a general framework for the conformal anomaly in interacting theories, using the framework of locally covariant field theory, as developed in \cite{HollandsWaldWick, HollandsWaldTO, HollandsWaldStress, HollandsYM}. We consider an action $S$ which we split as $S = S_0 + S_\ia$ into a free part $S_0$ (quadratic in the fields) and an interaction $S_\ia$ of higher order in the fields. 
In the following, we specialize the general framework introduced in the above references to the case of a free action $S_0$ which is invariant under local conformal transformations (typically, we are interested in interactions $S_\ia$ with the same property, but that is not necessary to set up our framework). 

We describe the framework of locally covariant field theory to the extent necessary for our purposes. For concreteness, we will consider a real scalar field, but the generalization to other fields (such as Dirac fields) is straightforward, \cf \cite{LocCovDirac} for example. 
To each globally hyperbolic four-dimensional spacetime $(M, g)$, one associates an algebra $( \F(M, g), \star, * )$ of functionals, with non-commutative product $\star$ and involution $*$. 
Concretely, we consider functionals
\beq
\label{eq:GeneralFunctional}
 F[\phi] = \sum_{k = 0}^N \int_{M^k} f_k(x_1, \ldots, x_k) \phi(x_1) \cdots \phi(x_k) \vol(x_1) \cdots \vol(x_k)
\eeq
with $f_k$ compactly supported symmetric distributions on $M^k$ whose singularities are restricted by a condition on their wave front set \cite{DuetschFredenhagenLoopExpansion, HollandsWaldWick}.
The \emph{support} of such a functional is defined as the union of the projections of the supports of $f_k$ from $M^k$ to $M$ for $k \geq 1$.
A functional independent of $\phi$ (i.e., $N=0$ in the above) is called a c-number functional. The involution $*$ acts by complex conjugation of $f_k$, while the $\star$ product is given by
\beq
 F \star G \defeq F \exp ( \hbar \, \cevvec{\Gamma}_{w} ) G \,,
\eeq
where
\beq
 F \, \cevvec \Gamma_w G = \int_{M^2} \frac{\delta F}{\delta \phi(x)} w(x, x') \frac{\delta G}{\delta \phi(x')} \vol(x) \vol(x') \,,
\eeq
and $w(x, x')$ is a Wightman two-point function of Hadamard form \cite{Radzikowski}. The $\star$ product depends on the choice of the two-point function $w$, but the algebras for different choices are canonically $*$-isomorphic \cite{HollandsWaldWick}. 
The c-number functionals are multiples of the identity \wrt the $\star$ product.

In order to later make sense of perturbative expansions, it is useful to introduce a grading on $( \F(M, g), \star, * )$, namely
\beq
 \Deg \defeq \deg_\phi + 2 \deg_\hbar \,,
\eeq
where $\deg_\phi$ counts the number of fields (a functional of the form \eqref{eq:GeneralFunctional} with $f_k = 0$ for $k \neq n$ and $f_n \neq 0$ would have $\deg_\phi = n$) and $\deg_\hbar$ the power in $\hbar$. The $\star$ product is additive \wrt this grading. 

To implement dynamics, one divides out the $*$ ideal $\mathfrak{I}(M, g)$ generated by the free equations of motion, i.e., for the conformally coupled scalar field, the functionals of the form\footnote{To bring this into the form \eqref{eq:GeneralFunctional}, one uses integration by parts to let $\Box - \frac{1}{6} R$ act on the first variable of $f_k$ and then symmetrizes.}
\beq
 \sum_{k = 1}^N \int_{M^k} f_k(x_1, \dots, x_k) \left( \Box - \tfrac{1}{6} R \right) \phi(x_1) \phi(x_2) \cdots \phi(x_k) \vol(x_1) \cdots \vol(x_k) \,.
\eeq
The quotient algebra is called the \emph{on-shell algebra} $\F^\os(M, g)$, and equality in the on-shell algebra is denoted by $\simeq$.

There are three cases in which (on-shell) algebras for different background spacetimes $(M, g)$ are related by $*$-homomorphisms, or even $*$-isomorphisms. 
These are relevant for formulating important constraints on time-ordered products below. The most straightforward case occurs when $(M, g)$ is a sub-spacetime of $(M', g')$, or, more generally, when there is an isometric embedding $\chi \colon (M, g) \to (M', g')$ preserving orientation, time-orientation, and causality.\footnote{$\chi$ preserves causality if every causal curve in $M'$ connecting two points in $\chi(M) \subset M'$ is contained in $\chi(M)$.} Then we define $\alpha_\chi \colon \F^{(\os)}(M, g) \to \F^{(\os)}(M', g')$ by
\beq
 (\alpha_\chi F)[\phi] \defeq F[ \chi^* \phi ] \,,
\eeq
i.e., by the pullback of $\phi$ along $\chi$. Equivalently, one can also define it via the push-forward along $\chi^{\otimes k}$ of the compactly supported distributions $f_k$ of \eqref{eq:GeneralFunctional} (with $\chi^{\otimes k}$ the obvious extension of $\chi \colon M \to M'$ to $\chi^{\otimes k} \colon M^k \to M^{\prime k}$). 
When the $\star$-products of $\F^{(\os)}(M, g)$ and $\F^{(\os)}(M', g')$ are defined with compatible two-point functions $w$, $w'$, in the sense that $w = (\chi^{\otimes 2})^* w'$, then $\alpha_\chi$ is a $*$-homomorphism (and even a $*$-isomorphism when $\chi$ is an isometry, i.e., an isometric diffeomorphism). 
Since, as mentioned above, algebras defined \wrt different two-point functions are canonically $*$-isomorphic, we may, given $w'$, without loss of generality simply choose $w$ in that manner.

A further isomorphism between algebras defined on different spacetimes occurs when the metric $g$ is conformally rescaled, i.e., between $\F(M, g_{\mu \nu})$ and $\F(M, \Omega^2 g_{\mu \nu})$ for some smooth positive function $\Omega$ on $M$. Given such an $\Omega$, one defines $\gamma_\Omega \colon \F^{(\os)}(M, \Omega^2 g) \to \F^{(\os)}(M, g)$ by
\begin{equation}
 (\gamma_\Omega F)[\phi] \defeq F[ \Omega^{-1} \phi ] \,.
\end{equation}
When the $\star$-products of $\F^{(\os)}(M, g)$ and $\F^{(\os)}(M, g^\Omega = \Omega^2 g)$ are defined with compatible two-point functions $w$, $w^\Omega$, in the sense that 
\beq
\label{eq:W_Omega}
 w^\Omega(x, x') = \Omega^{-1}(x) \Omega^{-1}(x') w(x,x') \,,
\eeq
then $\gamma_\Omega$ is a $*$-isomorphism. 
A special case that we will be considering is a constant scaling factor, which will be denoted by $\eta$. 
In fact, in this case the $*$-isomorphism $\gamma_\eta$ can be defined also for general (not necessarily conformal) theories, when all dimensionful parameters of the free theory (such as a mass) are scaled accordingly. This is the scaling transformation that is actually considered in the general framework of locally covariant field theory \cite{HollandsWaldWick, HollandsWaldTO}, where conformal invariance of the free action is not required.\footnote{An alternative possibility to incorporate local conformal invariance into the framework of locally covariant field theory was pursued in \cite{PinamontiConformal}, namely generalizing $\alpha_\chi$ to encompass also conformal embeddings $\chi$.}

Finally, we consider the situation where a metric $g'_{\mu \nu}$ is obtained by a compactly supported variation of $g_{\mu \nu}$, i.e., $g_{\mu \nu}$ and $g'_{\mu \nu}$ coincide except on a compact subset of $M$. We can then define the \emph{retarded variation} on field configurations by \cite{BreDue}
\beq
\label{eq:tauret_def}
 \tau^\ret_{g, g'} \phi \defeq \phi + E^\ret ( (P' - P) \phi ) \,,
\eeq
with $P$, $P'$ the Klein-Gordon operators $\Box - \frac{1}{6} R$ \wrt $g$, $g'$, and $E^\ret$ the retarded propagator \wrt $P$. Obviously $\tau^\ret$ is linear, $\tau^\ret_{g, g'} \phi$ coincides with $\phi$ except on the causal future of the support of $g_{\mu \nu} - g'_{\mu \nu}$, and $P \tau^\ret_{g, g'} \phi = P' \phi$, so when $\phi$ solves $P' \phi = 0$, then $\tau^\ret_{g, g'} \phi$ solves $P \tau^\ret_{g, g'} \phi = 0$. One then defines $\tau^\ret_{g, g'} \colon \F^{(\os)}(M, g') \to \F^{(\os)}(M, g)$ by (note the change of the order of $g$ and $g'$ on the two sides of the equation)
\beq
 (\tau^\ret_{g,g'} F)[\phi] \defeq F[ \tau^\ret_{g', g} \phi ] \,.
\eeq
When the $\star$-products of $\F^{(\os)}(M, g)$ and $\F^{(\os)}(M, g')$ are defined with compatible two-point functions $w$, $w'$, in the sense that (here the tensor product of $\tau^\ret_{g',g}$ indicates that the map acts on both variables as in \eqref{eq:tauret_def})
\beq
 w' = \tau^\ret_{g',g} \otimes \tau^\ret_{g',g} w \,,
\eeq
then $\tau^\ret_{g, g'}$ is a $*$-isomorphism. 

When $\Omega$ is such that $\Omega = 1$ except on a compact subset of $M$, then we can compare $\gamma_\Omega$ and $\tau^\ret_{g, \Omega^2 g}$, which are both maps $\F^{(\os)}(M, \Omega^2 g) \to \F^{(\os)}(M, g)$. 
For a solution $\phi$ to the Klein-Gordon equation $P \phi = 0$, we have $\tau^\ret_{\Omega^2 g, g} \phi = \Omega^{-1} \phi$, which follows from the fact that both satisfy the Klein-Gordon equation \wrt the scaled metric $\Omega^2 g_{\mu \nu}$ and coincide in a neighborhood of a Cauchy surface to the past of the region where $\Omega \neq 1$. 
It follows that, on the on-shell algebras, $\gamma_\Omega$ and $\tau^\ret_{g, \Omega^2 g}$ coincide, i.e.,\footnote{The analogous property holds for $\alpha_\chi$ and $\tau^\ret_{g, g'}$ when $\chi$ is an isometry $\chi \colon (M, g) \to (M, g')$ which coincides with the identity on the complement of a compact subset \cite[Prop.~2.10]{BackgroundIndependence}.}
\beq
\label{eq:gamma_Omega_simeq_tau_ret}
 \gamma_\Omega F \simeq \tau^\ret_{g, \Omega^2 g} F \,.
\eeq

Later, we will employ infinitesimal versions of $\gamma_\Omega$ and $\tau^\ret_{g, g'}$. 
For that, assume that we have a family $F_{g} \in \F(M, g)$ of functionals depending smoothly\footnote{The smoothness requirement can be made precise as follows: For any smooth $m$ parameter family $g^{(s)}_{\mu \nu}$ of metrics (with $s$ varying over an open subset $U$ of $\R^m$), we require that the corresponding ``kernels'' $\tilde f_k(s, x_1, \dots, x_k)$, \cf \eqref{eq:GeneralFunctional}, now seen as distributions on $U \times M^k$, fulfill an appropriate wave front set condition. Specifically, we require that, for each $k$, the wave front set of $\tilde f_k$ is contained in $\{ (s, \rho; x_1, p_1; \dots x_k, p_k) \in T^* (U \times M^k) \setminus \{ 0 \} | (x_1, p_1; \dots x_k, p_k) \subset W^k_{g^{(s)}} \}$, with $W^k_g$ the subset of $T^* M^k \setminus \{ 0 \}$ to which the wave front set of $f_k$ is restricted to in order for a functional $F$ (as given by \eqref{eq:GeneralFunctional}) to be an element of $\F(M, g)$, \cf \cite{DuetschFredenhagenLoopExpansion, HollandsWaldWick}. This condition in particular ensures that an arbitrary number of functional derivatives \wrt the metric can be taken.} on the background metric $g$. We then define, for an infinitesimal variation $f$ of the scale factor or $h_{\mu \nu}$ of the metric,
\begin{align}
\label{eq:delta_s_f}
 \delta^\weyl_f F & \defeq \del_\epsilon \left( \gamma_{1 + \epsilon f} F_{(1 + \epsilon f)^2 g} \right) \big\rvert_{\epsilon = 0} \,, &
 \delta^\ret_h F & \defeq \del_\epsilon \left( \tau^\ret_{g, g+ \epsilon h} F_{g + \epsilon h} \right) \big\rvert_{\epsilon = 0} \,,
\end{align}
namely $\delta^\weyl$ is an infinitesimal Weyl transformation and $\delta^\ret$ an infinitesimal retarded variation. As derivatives of $*$-isomorphisms, these are real derivatives, i.e., they fulfill the Leibniz rule \wrt the $\star$ product. Furthermore, $\delta^\ret_{2 f g}$ coincides on-shell with $\delta^\weyl_f$, by \eqref{eq:gamma_Omega_simeq_tau_ret}.

A special class of functionals within $\F(M, g)$ (a subspace, but not a subalgebra) are the \emph{local functionals} $\F_\loc(M, g)$, which can be written in the form
\beq
 F[\phi] = \sum_{k = 0}^N \int_M \nabla_{\alpha_1} \phi(x) \dots \nabla_{\alpha_k} \phi(x) f_k^{\alpha_1 \dots \alpha_k}(x) \vol(x) \,,
\eeq
with $\alpha_i$ multi-indices and $f_k^{\alpha_1 \dots \alpha_k}$ smooth compactly supported tensors. 

All local functionals can be written as linear combinations of smeared fields. Here a \emph{field} $\Phi$ associates to any spacetime $(M, g)$ a linear map $\Phi_{(M, g)} \colon \Gamma(M, T^\otimes M) \to \F_\loc(M, g)$ such that for any isometric embedding $\chi \colon (M, g) \to (M', g')$ as discussed above
\beq
 \alpha_\chi \Phi_{(M, g)}(t) = \Phi_{(M', g')}( \chi_* t) \,,
\eeq
where $t$ is a smooth compactly supported tensor of the appropriate index structure and $\chi_* t$ is its pushforward. 
Examples are $(\nabla_{\alpha_1} \phi \dots \nabla_{\alpha_k}\phi)(x) \defeq \nabla_{\alpha_1} \phi(x) \dots \nabla_{\alpha_k} \phi(x)$, but there are also c-number fields constructed out of $g_{\mu \nu}$, $g^{\mu \nu}$, $\tensor{R}{_{\mu \nu \lambda}^\rho}$ and its covariant derivatives.\footnote{c-number fields are not needed to generate local functionals, but they give rise to families of local (c-number) functionals for different backgrounds. They also naturally occur as anomalies, \cf the example of $\phi^4$ theory studied in Section~\ref{eq:phi4_Cohomology}.}
One defines the \emph{scaling dimension} of a field $\Phi$ as the number $d_\Phi$ such that for a constant scaling factor $\eta$
\beq
 \gamma_\eta \Phi_{(M, \eta^2 g)}(\eta^{d_\Phi - 4} t) = \Phi_{(M, g)}(t) \,.
\eeq
One easily checks that the field $\nabla_{\alpha_1} \phi \dots \nabla_{\alpha_k}\phi$ from above has scaling dimension $k$. 
As $g_{\mu \nu}$ and $g^{\mu \nu}$ scale (with scaling dimensions $2$ and $-2$), the scaling dimension depends on the index position. 
The \emph{mass dimension}, which is defined as the scaling dimension plus the number of lower minus the number of upper indices, is independent of the index position and coincides with the usual notion of the mass dimension (in particular, each derivative increases the mass dimension by one and each curvature $R$, irrespective of indices, by two).

In order to define interacting observables, one uses \emph{time-ordered products}. In the present context, these are linear maps $\cT_{(M, g)} \colon \F_\loc^{\otimes k}(M, g) \to \F(M, g)$ from tensor products of local functionals into the observable algebra $\F(M, g)$, fulfilling certain requirements. The time ordered products with a single factor, as occurring in \eqref{eq:FreeTraceAnomalyGeneral} above, are also called \emph{Wick powers}. 
We already mentioned that local functionals can be expressed in terms of the fields $\nabla_{\alpha_1} \phi \dots \nabla_{\alpha_k} \phi$ integrated with appropriate test tensors. 
It is thus sometimes useful to express time-ordered products in a distributional notation in the form
\beq
 \cT( (\nabla_{\alpha_1} \phi \dots \nabla_{\alpha_k} \phi) (x_1) \otimes (\nabla_{\beta_1} \phi \dots \nabla_{\beta_l} \phi)(x_2) \otimes \dots ) \,,
\eeq
which still needs to be integrated with appropriate test tensors in order to give an element of $\F(M,g)$. 
However, the expression of a local functional in terms of smeared fields of the form $\nabla_{\alpha_1} \phi \dots \nabla_{\alpha_k} \phi$ is not unique: Considering for simplicity a functional quadratic in the fields, we have
\begin{multline}
 \int_M \nabla_{\mu_1} \dots \nabla_{\mu_s} \phi \nabla_{\nu_1} \dots \nabla_{\nu_t} \phi \nabla_\mu t^{\mu \mu_1 \dots \mu_s \nu_1 \dots \nu_t} \vol = \\ - \int_M \left( \nabla_\mu \nabla_{\mu_1} \dots \nabla_{\mu_s} \phi \nabla_{\nu_1} \dots \nabla_{\nu_t} \phi + \nabla_{\mu_1} \dots \nabla_{\mu_s} \phi \nabla_\mu \nabla_{\nu_1} \dots \nabla_{\nu_t} \phi \right) t^{\mu \mu_1 \dots \mu_s \nu_1 \dots \nu_t} \vol \,.
\end{multline}
While on the \lhs we have a field involving in total $s + t$ derivatives, we have fields with in total $s + t + 1$ derivatives on the right hand side. One says that the combination of fields occurring on the \rhs is Leibniz dependent. Such relations must be respected when expressing time-ordered products in terms of fields (as we will do in our concrete calculations below). As an elementary example, we must have
\beq
\label{eq:nabla_T_phi_2}
 \nabla_\mu \cT( \phi^2(x) \otimes F_1 \otimes \dots ) = 2 \cT ( (\phi \nabla_\mu \phi)(x) \otimes F_1 \otimes \dots ) \,,
\eeq
so that the time-ordered products with a factor of $\phi \nabla_\mu \phi$ (which is Leibniz dependent) are completely determined by those with a factor of $\phi^2$.

Time-ordered products fulfill a set of axioms \cite{HollandsWaldWick, HollandsWaldTO, HollandsWaldStress}:

\begin{description}
\item[Symmetry:] Time-ordered products are symmetric in the tensor factors, i.e.,
\beq
\label{eq:TO_Symmetry}
 \cT( F_1 \otimes \dots F_i \otimes \dots F_j \otimes \dots F_k ) = \cT( F_1 \otimes \dots F_j \otimes \dots F_i \otimes \dots F_k ) \,.
\eeq

\item[$\hbar$ expansion:] Time-ordered products respect the $\Deg$ grading in the sense that
\beq
\label{eq:TO_Deg}
 \Deg \cT(F_1 \otimes \dots \otimes F_k) = \sum_{i=1}^k \Deg F_i \,.
\eeq
\end{description}

Due to these properties, it is possible to express many of the following properties in terms of the generating functional\footnote{When functionals $F_i$ can be bosonic or fermionic, the symmetry requirement \eqref{eq:TO_Symmetry} is generalized to graded symmetry. For a fermionic functional $F$, the generating functional notation can be used by multiplying $F$ with a formal fermionic parameter (to have a bosonic exponent).} (to be understood as a formal series in $F$)
\beq
 \cT( \et{F} ) = \sum_{k=0}^\infty \frac{1}{k!} \cT ( \underbrace{ F \otimes \dots \otimes F }_{k \text{ times }} ) \,.
\eeq

\begin{description}

\item[Linear field:] Time-ordered products involving a linear field 
satisfy
\beq
 \cT( \phi(x) \otimes \et{F} ) = \phi(x) \star \cT( \et{F} ) + \rmi \hbar \int_M \cT ( \tfrac{\delta}{\delta \phi(x')} F \otimes \et{F} ) E^\adv(x, x') \vol(x') \,,
\eeq 
with $E^\adv$ the advanced propagator associated to the Klein-Gordon operator $\Box - \frac{1}{6} R$.

\item[Field independence:] Time-ordered products commute with functional differentiation in the sense that
\beq
\label{eq:TO_FieldIndependence}
 \tfrac{\delta}{\delta \phi(x)} \cT ( \et{F} ) = \cT ( \tfrac{\delta}{\delta \phi(x)} F \otimes \et{F} ) \,.
\eeq

\end{description}

The linear field axiom implements the notion that a linear field does not require any renormalization, so a time-ordered product involving such should be expressible in terms of time-ordered products of fewer factors. The axiom guarantees that, in the appropriate sense, the interacting field fulfills the interacting equations of motion \cite{HollandsWaldStress}. The field independence axiom, formulated here as in \cite{BDF09}, is the analog of the possibility to perform integration by parts in a formal path integral.

One consequence of these two axioms is that for a functional $F_0$ which is independent of $\phi$, i.e., a c-number functional, we have\footnote{For example, as any c-number functional $F_0$ can be expressed as $F_0 = \int \frac{\delta}{\delta \phi(x)} \phi(f) h(x) \vol(x)$ for suitably chosen $f$, $h$, the linear field and field independence axioms imply that $\cT(F_0) = F_0$.}
\beq
 \cT(F_0 \otimes \dots \otimes F_k) = F_0 \cT(F_1 \otimes \dots \otimes F_k) \,.
\eeq
Furthermore, for a Wick power $\cT(F)$, field independence implies that $\cT(F)$ is also a local functional.\footnote{The second derivative $\frac{\delta^2}{\delta \phi(x) \delta \phi(x')} \cT(F)$ is supported at coinciding points $x = x'$, on account of $F$ being a local functional.} The linear field axiom implies that $\cT$ acts trivially on a linear field, $\cT( \phi(f) ) = \phi(f)$. Again from \eqref{eq:TO_FieldIndependence} and \eqref{eq:TO_Deg} it follows that $\cT(F) = F + \cO(\hbar)$, with the higher order corrections of lower order in the fields. In particular, it follows that, on local functionals, $\cT$ can be inverted (in the sense of a formal power series in $\hbar$), i.e., we have $\cT^{-1} \colon \F_\loc \to \F_\loc$.

The next two axioms encode the time-ordering property and ensure the unitarity of the $S$ matrix constructed out of time-ordered products.

\begin{description}

\item[Causal factorization:] Time-ordered products factorize when the arguments are in causal order, i.e.
\beq
 \cT( \et{F} \otimes \et{G} ) = \cT (\et{F} ) \star \cT( \et{G} )
\eeq 
whenever $\supp F \cap J^-(\supp G) = \emptyset$ (meaning that $\supp F$ is later than $\supp G$ \wrt to some Cauchy surface separating the two).

\item[Unitarity:] Time-ordered products are unitary in the sense that
\beq
 \cT ( \et{ \rmi F} )^* \star \cT( \et{ \rmi F^* } ) = 1 \,.
\eeq

\end{description}

The causal factorization axiom allows to recursively define time-ordered products by extending distributions defined up to coinciding points, i.e., on $M^k \setminus \{ (x, \dots, x) \mid x \in M \}$, to distributions defined everywhere, i.e., on $M^k$ \cite{EpsteinGlaser, BrunettiFredenhagenScalingDegree, HollandsWaldTO}. In order to enforce that this is done in a coherent manner for different spacetime backgrounds $(M, g)$, one requires some relations between time-ordered products on different spacetimes. Above, we already introduced certain $*$-isomorphisms (or $*$-homomorphisms) between the algebras $\F(M, g)$, $\F(M', g')$ for different spacetimes. Using these, we require the following:

\begin{description}

\item[Local covariance:] For $\chi \colon (M, g) \to (M', g')$ an isometric embedding preserving orientation, time-orientation, and causality, we have
\beq
 \alpha_\chi \cT_{(M, g)}( \et{F} ) = \cT_{(M', g')}( \et{ \alpha_\chi F} ) \,.
\eeq

\item[Almost homogeneous scaling:] Time-ordered products scale almost homogeneously in the sense that for each collection $\Phi_1, \dots, \Phi_k$ of homogeneously scaling fields, there is a finite number $l$ such that
\beq
\label{eq:TO_Scaling}
 \left( \eta \del_\eta \right)^l \gamma_{\eta} \cT_{(M, \eta^2 g)}( \Phi_{1, (M, \eta^2 g)} ( \eta^{ d_{\Phi_1} - 4} t_1 ) \otimes \dots \otimes \Phi_{k, (M, \eta^2 g)} ( \eta^{ d_{\Phi_k} - 4} t_k ) ) = 0 \,.
\eeq

\item[Perturbative agreement:] For any compactly supported infinitesimal variation $h_{\mu \nu}$ of the metric, we have 
(\cf \eqref{eq:RetardedProduct} for the definition of $\cR$ and \eqref{eq:delta_h_F} for the definition of $\delta_h$)
\beq
\label{eq:TO_PA}
 \delta^\ret_h \cT ( \et{F} ) = \cT( \delta_h F \otimes \et{F} ) + \ibar \cR ( \et{F}; \delta_h S_0 ) \,.
\eeq

\end{description}

A few comments and explanations on these requirements are in order: While local covariance ensures that the renormalization (extension of distributions) necessary to define time-ordered products is done in a local and covariant manner, almost homogeneous scaling ensures that the usual power counting rules are respected. Regarding the formulation of the latter, we note that homogeneous scaling of time-ordered products would amount to
\begin{multline}
\label{eq:TO_HomogeneousScaling}
 \gamma_{\eta} \cT_{(M, \eta^2 g)}( \Phi_{1, (M, \eta^2 g)} ( \eta^{ d_{\Phi_1} - 4} t_1 ) \otimes \dots \otimes \Phi_{k, (M, \eta^2 g)} ( \eta^{ d_{\Phi_k} - 4} t_k ) ) \\
 = \cT_{(M, g)}( \Phi_{1, (M, g)} ( t_1 ) \otimes \dots \otimes \Phi_{k, (M, g)} ( t_k ) ) \,,
\end{multline}
in which case \eqref{eq:TO_Scaling} would be fulfilled for $l = 1$. If \eqref{eq:TO_Scaling} does not hold for $l < L$, but for $l = L$, then the \rhs of \eqref{eq:TO_HomogeneousScaling} receives corrections which are polynomials in $\ln \eta$ of order $L-1$. 
These logarithmic corrections in general originate from i) the almost homogeneous scaling behavior of the Hadamard parametrix, which is used to construct Wick powers (time-ordered products of a single factor) and ii) the extension process discussed above which in general turns an originally homogeneous (but not everywhere defined) distribution into an almost homogeneous one.

As for perturbative agreement, we still need to explain some of the symbols on the \rhs of \eqref{eq:TO_PA}. The \emph{retarded product} $\cR$ is defined by the generating functional
\beq
\label{eq:RetardedProduct}
 \cR(\et{F} ; \et{G}) = \cT( \et{G} )^{-1} \star \cT( \et{F} \otimes \et{G} ) \,,
\eeq
with the inverse (in the sense of formal power series in $G$) \wrt the $\star$ product. In particular $\cR( \et{F} ; \et{G} ) = \cT( \et{F} )$ whenever $J^-(\supp F) \cap \supp G = \emptyset$. Supplied with appropriate powers of $\ibar$, this is the generating functional for time-ordered products of observables $F$ in the presence of an interaction $G$. It will thus be used to define (time-ordered products of) interacting observables below. The second as yet unexplained notation in \eqref{eq:TO_PA} is $\delta_h F$, which stands for the functional derivative of the local functional $F$ \wrt the metric $g_{\mu \nu}$ in the direction $h_{\mu \nu}$, i.e.,
\beq
\label{eq:delta_h_F}
 \delta_h F[g, \phi] \defeq \del_\epsilon F[g+ \epsilon h, \phi] \big\rvert_{\epsilon = 0} \,.
\eeq
When we introduce the stress tensor below in \eqref{eq:StressTensor}, we will see that $\delta_h S_0$ is actually the free part $T^{0,\mu \nu}$ of the stress tensor, integrated against $\frac{1}{2} h_{\mu \nu}$. 
Perturbative agreement (in particular in its application to interacting time-ordered products defined below) expresses the notion that it should not matter whether one implements a local change $g \to g'$ of the metric by a change of the background metric or by including the difference $S[g', \phi] - S[g, \phi]$ in the interaction. It corresponds to the ``renormalized action principle'' (or ``quantum action principle'') \cite{Lowenstein71, Lam72, BreitenlohnerMaison} used in \cite{BrownCollins, HathrellScalar}. Perturbative agreement can be fulfilled whenever there is a definition of Wick powers (time-ordered products of a single factor) such that $\cT ( \nabla^\mu T^0_{\mu \nu} ) \simeq 0$, i.e., the divergence of the Wick power of the free current vanishes in the on-shell algebra \cite{HollandsWaldStress}.

Apart from the above axioms, one further requires a certain regularity of time-ordered products, for which there are several alternative formulations \cite{HollandsWaldWick, HollandsWaldTO, KhavkineMoretti, MoroPhD, HofmannPhD}, but which we will not elaborate upon.

For the concrete calculations that we are going to perform in our analysis of the $\phi^4$ theory, we will have to explicitly construct time-ordered products for up to two factors. For this reason, we now explain the basic idea for their construction. Wick products (time-ordered products of a single factor) fulfilling all axioms except for perturbative agreement can be defined by the Hadamard point-split prescription. If $w$ is the two-point function used to define the $\star$-product, then one defines
\beq
\label{eq:HadamardPointSplit}
 \cT( F ) = \exp( \hbar \Gamma_{w - H} ) F
\eeq
with
\beq
 \Gamma_D = \frac{1}{2} \int_{M^2} D(x, x') \frac{\delta^2}{\delta \phi(x) \delta \phi(x')} \vol(x) \vol(x') \,.
\eeq
Here $H(x, x')$ is the \emph{Hadamard parametrix} which is defined in a neighborhood of coinciding points, is constructed locally and covariantly, and captures the singularities of the Hadamard two-point function $w$ (so that $w-H$ is smooth). With this definition, the stress tensor is not conserved on-shell, but for typical field theories\footnote{Counterexamples would be theories with gravitational anomalies, such as chiral fermions in $4k +2$ spacetime dimensions \cite{AlvarezGaumeWitten}.} (in particular the scalar field in four spacetime dimensions), this (and thus also perturbative agreement) can be achieved by exploiting the remaining renormalization freedom \cite{HollandsWaldStress}.

Proper time-ordered products (with more than one factor), can be first defined formally via
\beq
\label{eq:TO_Formal}
 \cT(F_1 \otimes \dots \otimes F_k) = \exp \left( \hbar \sum_{i < j} \Gamma^{ij}_{w_\feyn} \right) \cT(F_1) \dots \cT(F_k) \,.
\eeq
Here
\beq
 \Gamma^{ij}_{w_\feyn} \defeq \int_{M^2} w_\feyn(x,x') \frac{\delta^i}{\delta \phi(x)} \frac{\delta^j}{\delta \phi(x')} \vol(x) \vol(x')
\eeq
with $\frac{\delta^i}{\delta \phi(x)}$ acting on $\cT(F_i)$, and $w_\feyn$ being the \emph{Feynman propagator}
\beq
 w_\feyn(x, x') = w(x, x') + \rmi E^\adv(x, x')
\eeq
associated to the two-point function $w(x,x')$ defining the $\star$ product. When the local functionals $F_i$ are of higher than linear order in the field, then the above leads to products of Feynman propagators which are only well-defined as distributions up to coinciding points. The definition \eqref{eq:TO_Formal} turns out to be well-defined when the supports of the $F_i$ do not overlap. As already indicated above, to fully define time-ordered products, one then has to extend certain (products of) distributions to $M^k$. Below, we will concretely perform this for up to the fourth power of the Feynman propagator $w_\feyn$, which otherwise would only be defined on $M^2 \setminus \{ (x, x) \mid x \in M \}$. For a proof that this extension is possible in general such that the above axioms are fulfilled, we refer to \cite{HollandsWaldTO} (or \cite{HollandsWaldStress} for the inclusion of perturbative agreement).

The above axioms do not fix time-ordered products uniquely. The remaining ambiguity is encoded in the \emph{main theorem of renormalization} \cite{PopineauStora, HollandsWaldTO, DuetschFredenhagenAWI, HollandsWaldRG}: Two schemes $\cT$, $\tilde \cT$ fulfilling the above axioms are related by
\beq
\label{eq:MainTheorem}
 \tilde \cT(\eti{F}) = \cT( \eti{(F + Z(\et{F}))} ) \,,
\eeq
with linear maps $Z \colon \F_\loc^{\otimes k} \to \F_\loc$ which are symmetric, at least of $\cO(\hbar)$, and fulfill (the correction term on the \rhs is due to the manner in which $\hbar$ is included in the exponent in \eqref{eq:MainTheorem}) 
\beq
\label{eq:Z_F}
 \Deg Z( F_1 \otimes \dots \otimes F_k ) = \sum_{i = 1}^k \Deg F_i - 2 (k-1) \,.
\eeq
Furthermore, $Z$ is field independent in exactly the same manner as $\cT$, \cf \eqref{eq:TO_FieldIndependence}, and vanishes if one of the factors is a linear field or a c-number. $Z(F_1 \otimes \dots \otimes F_k)$ vanishes unless the supports of all $F_i$ overlap and has support contained in this overlap. It is real in the sense that $Z(\et{F})^* = Z(\et{F^*})$. 
It is also locally covariant and scales homogeneously, i.e., \eqref{eq:TO_Scaling} holds for $Z$ with $l = 1$. $Z$ depends analytically on the background geometry in the sense that for fields $\Phi_1, \dots, \Phi_k$,
\beq
\label{eq:Z_Phi_1_Phi_k}
 Z( \Phi_1(t_1) \otimes \dots \otimes \Phi_k(t_k) ) = \sum_j \Psi_j(s_j) \,,
\eeq
where $\Psi_j$ are some other fields and the test tensors $s_j$ are constructed as the product of covariant derivatives of the $t_i$ multiplied by a polynomial in $g_{\mu \nu}$, $g^{\mu \nu}$ and covariant derivatives of $\tensor{R}{_{\mu \nu \lambda}^\rho}$.\footnote{The proof of this statement depends on the regularity conditions imposed on time-ordered products, and we are only aware of a proof under the strong regularity assumptions used in \cite{HollandsWaldTO}.}
Of course, being defined on local functionals, $Z$ must respect Leibniz dependencies as discussed above for time-ordered products, for example the relation \eqref{eq:nabla_T_phi_2} also holds for $Z$. Finally, in order to preserve perturbative agreement in the redefinition, one must have \cite{AdlerBardeenThm} (\cf \cite{Habil} for a proof)
\beq
\label{eq:delta_ret_h_Z}
 \delta_h Z(\et{F}) = Z( \{ \delta_h F + \delta_h S_0 \} \otimes \et{F} ) - Z (\delta_h S_0) \,.
\eeq
Conversely, given time-ordered products $\cT$ and maps $Z$ fulfilling the above properties, one can define a new set $\tilde \cT$ of time-ordered products by \eqref{eq:MainTheorem}.

A particular example for a redefinition of time-ordered products is given by a scale transformation. Namely, given time-ordered products $\cT$, we can define, for any $\eta > 0$, new time-ordered products $\cT^{(\eta)}$ by
\beq
\label{eq:T_eta}
 \cT^{(\eta)}_{(M, g)}( \eti{ \Phi_{(M, g)}(t) } ) \defeq \gamma_\eta \cT_{(M, \eta^2 g)} ( \eti{ \Phi_{(M, \eta^2 g)}( \eta^{d_\Phi-4} t)} ) \,.
\eeq
As local functionals can be expressed in terms of fields, this defines a new set of time-ordered products. By the main theorem of renormalization, we must have
\beq
\label{eq:Z_eta}
 \cT^{(\eta)}(\eti{F}) = \cT( \eti{(F + Z^{(\eta)}(\et{F}))})
\eeq
for a map $Z^{(\eta)}$ with the above properties, which is a polynomial in $\ln \eta$. This is the basis for the definition of the renormalization group in locally covariant field theory \cite{HollandsWaldRG}.

As already indicated, (time-ordered products of) interacting observables are constructed using the retarded product \eqref{eq:RetardedProduct}. For this, we need to localize the interaction, i.e., introduce an infrared cutoff function, which we typically denote by $\chi$. To be specific, $\chi(x)$ is smooth, compactly supported, and equal to $1$ in a neighborhood of a (causally convex) spacetime region $\cO$ (the region within which we want to consider local observables). The interacting part of the action is then obtained by integrating the interacting part $L_\ia$ of the Lagrangian with $\chi$, i.e.,
\beq
\label{eq:S_int_chi}
 S_\ia(\chi) = \int_M \chi L_\ia \vol \,.
\eeq
As explained below, the precise form of $\chi$ does not matter, and we will typically simply write $S_\ia$.
The generating functional of interacting time-ordered products of local functionals with support contained in $\cO$ is then given by \emph{Bogoliubov's formula} \cite{IlinSlavnov}
\beq
\label{eq:BogoliubovsFormula}
 \cT^\ia (\eti{F}) \defeq \cR( \eti{F} ; \eti{ S_\ia }) \,,
\eeq
where $\supp F \subset \cO$. The algebra $\F_\ia(\cO)$ generated by these depends on the choice of the cutoff functions $\chi$, but in an inessential way: The algebras obtained by different choices of the cutoff are related by a unitary transformation \cite{BrunettiFredenhagenScalingDegree}. One can use this to define a global interacting algebra $\F_\ia(M)$ via the \emph{algebraic adiabatic limit} \cite{BrunettiFredenhagenScalingDegree, HollandsWaldRG}. For our purposes, this will not be relevant, as we will always be interested in the behavior of the observables under local scale transformations. We can thus simply assume that $\chi$ is equal to $1$ on the support of the scale transformation. Nevertheless, the issue of the localization of the interaction term will resurface in our discussion of the possibility to achieve a trace anomaly without terms such as $R^2$.

\section{The Weyl and the trace anomaly}
\label{sec:WeylAnomaly}

Before focussing on the trace of the stress tensor (the trace anomaly), we first develop a general framework for anomalies related to local scale transformations, called Weyl anomalies in the following. As already indicated above, we assume that the free part $S_0$ of the action is invariant under local scale transformations. 
We consider the infinitesimal version $\delta^\weyl_f$ of the scaling transformation, defined in \eqref{eq:delta_s_f}. 
On a field, it acts as a derivation, with action
\begin{align}
 \delta^\weyl_f g_{\mu \nu} & = 2 f g_{\mu \nu} \,, &
 \delta^\weyl_f \phi & = - f \phi
\end{align}
on the elementary constituents of a field.\footnote{The scaling behaviour of $\phi$ is that of a scalar field in four spacetime dimensions. For different fields or dimensions, this has to be adjusted.} The action on the inverse metric, the Riemann tensor, and covariant derivatives follow from these. 
We say that time-ordered products $\cT$ respect local scale transformations if $\delta^\weyl_f$ and $\cT$ commute, i.e., $\delta^\weyl_f \cT(\et{F}) = \cT (\delta^\weyl_f F \otimes \et{F})$.
The Weyl anomaly captures the violation of this relation. 
Specifically, we implicitly define $A_f$ by
\beq
\label{eq:A_f}
 \delta^\weyl_f \cT(\eti{F}) = \ibar \cT( \{ \delta^\weyl_f F + A_f(\et{F}) \} \otimes \eti{F}) \,.
\eeq
That this equation is consistent, i.e., that there is a local functional $A_f(\et{F})$ such that this holds, can be shown in complete analogy to the treatment of gauge anomalies \cf \cite{HollandsYM, FrobBV}. 
This is sketched in the Appendix, where it is also shown that $A_f( F_1 \otimes \dots F_k)$ is local in the scale function $f$, i.e., supported on the intersection of $\supp f$ with $\cap_j \supp F_j$. 
With one caveat (to be discussed below), the anomaly maps $A_f \colon \F_\loc^{\otimes k} \to \F_\loc$ can be shown to fulfill the same properties as the redefinition maps $Z$, discussed below \eqref{eq:MainTheorem}. In particular, they are field independent and vanish if one of the factors is a linear field (which implies that they also vanish if one of the factors is a c-number). The proof proceeds in complete analogy to that of the analogous properties of the gauge anomaly \cite{BgInd}. The caveat mentioned above is that we are not aware of a proof of analytic dependence of the anomaly on the background geometry even for the case of the gauge anomaly. We expect that such a proof can be given in close analogy to the one of the analogous property of the redefinition maps $Z$, \cf \eqref{eq:Z_Phi_1_Phi_k}. However, this proof relies on the regularity condition for time-ordered products, for which there are several proposals \cite{HollandsWaldWick, HollandsWaldTO, KhavkineMoretti}, and which we did not elaborate upon. Hence, we limit ourselves to stating that we expect analytic dependence of the anomaly on the background geometry in the sense that for fields $\Phi_j$ and corresponding test tensors $t_j^{\alpha_j}$, we have
\beq
\label{eq:A_f_HomogeneousScaling}
 A_f(\Phi_1(t_1) \otimes \dots \Phi_k(t_k)) = \int \Psi_{\alpha_1 \dots \alpha_k}^{\beta_0 \dots \beta_k} \nabla_{\beta_0} f \nabla_{\beta_1} t_1^{\alpha_1} \dots \nabla_{\beta_k} t_k^{\alpha_k} \vol
\eeq
for some fields $\Psi_{\alpha_1 \dots \alpha_k}^{\beta_0 \dots \beta_k}$ constructed polynomially out of $g_{\mu \nu}$, $g^{\mu \nu}$ and (covariant derivatives of) $\phi$ and the Riemann curvature tensor.

In order to derive a consistency relation analogous to the case of gauge theories \cite{HollandsYM, RejznerFredenhagenQuantization, FrobBV}, it is convenient to promote $\delta^\weyl_f$ to a fermionic operator. 
As in \cite{BonoraEtAl83}, we thus introduce a fermionic non-dynamical\footnote{\label{ft:T_xi_Phi}Being non-dynamical means that no field equations are imposed on $\xi$ in the on-shell algebra, that it is graded commuting with all other fields \wrt the $\star$ product, and that time-ordering acts trivially in the sense that $\cT( \xi(x) \Phi(x) \otimes F_1 \otimes \dots) = \xi(x) \cT( \Phi(x) \otimes F_1 \otimes \dots)$.} ghost field $\xi$ of vanishing mass dimension and define the differential
\begin{align}
 \Xi g_{\mu \nu} & = 2 \xi g_{\mu \nu} \,, &
 \Xi \phi & = - \xi \phi \,, &
 \Xi \xi & = 0 \,.
\end{align}
We then define the Weyl anomaly $A(\et{F})$ by
\beq
\label{eq:WeylAnomaly}
 \Xi \cT(\eti{F}) = \ibar \cT( \{ \Xi F + A(\et{F}) \} \otimes \eti{F}) \,.
\eeq
$A(\et{F})$ has properties analogous to $A_f(\et{F})$, except that it increases the number of ghost fields by one. When $F$ does not contain the ghost field $\xi$, then the relation to the anomaly $A_f$ defined in \eqref{eq:A_f} is $A_f(\et{F}) = A(\et{F})\rvert_{\xi \to f}$. 

Due to the fermionic nature of $\xi$, $\Xi$ is nilpotent, and the anomaly $A$ is fermionic. From the nilpotency of $\Xi$, one then obtains, in complete analogy to Proposition~4 of \cite{HollandsYM}, the \emph{consistency condition}
\beq
 \Xi A(\et{F}) + A( \{ \Xi F + A(\et{F}) \} \otimes \et{F}) = 0 \,.
\eeq
For reasons to be explained below, we will mainly be interested in $A(\et{S_\ia})$. If the interaction is invariant under local scale transformations, i.e., $\Xi S_\ia = 0$, we get
\beq
\label{eq:ConsistencyCondition}
 \Xi A(\et{S_\ia}) = - A ( A(\et{S_\ia}) \otimes \et{S_\ia}) \,.
\eeq
Now consider the anomaly $A(\et{S_\ia})$ order by order in $\hbar$ and assume that the $\cO(\hbar^m)$ contribution $A^{(m)}(\et{S_\ia})$ is the first non-vanishing one. As the anomaly increases the power in $\hbar$ at least by one, the \rhs of the above is of $\cO(\hbar^{m+1})$. It follows that the component $A^{(m)}(\et{S_\ia})$ of lowest non-vanishing order in $\hbar$ is $\Xi$ closed, which is an important structural constraint on the anomaly. Furthermore, it follows from the behaviour of $A$ under scaling that if $S_\ia$ is strictly renormalizable, i.e., the integral over a Lagrangian of mass dimension four, then so is $A( \et{S_\ia} )$.

For later purposes, it is important to note that under a redefinition \eqref{eq:MainTheorem} of time-ordered products generated by $Z$, the Weyl anomaly transforms in exactly the same manner as given in \cite[Prop.~21]{deMedeirosHollands} (in the context of superconformal gauge theory)
\beq
\label{eq:Xi_Z}
 \Xi Z(\et{F}) + A( \et{F + Z(\et{F})}) = Z( \{ \Xi F + \tilde A(\et{F}) \} \otimes \et{F} ) + \tilde A(\et{F}) \,,
\eeq
where $\tilde A$ is the anomaly for the redefined time-ordered products $\tilde \cT$. 
In particular, assuming that $\cO(\hbar^m)$ is the lowest non-vanishing order of the anomaly $A(\et{S_\ia})$, and that it is cohomologically trivial at this order,
\beq
 A^{(m)}(\et{S_\ia}) = \Xi F
\eeq
for a local functional $F$, then this anomaly can be removed at $\cO(\hbar^m)$ by the redefinition generated by
\beq
\label{eq:Z_exp_S_int}
 Z( \et{S_\ia}) = - F \,.
\eeq
For this reason, it is relevant to investigate the cohomology of $\Xi$.

Comparing the definition \eqref{eq:WeylAnomaly} of the Weyl anomaly with the definition \eqref{eq:T_eta} of $\cT^{(\eta)}$ and the corresponding redefinition maps $Z^{(\eta)}$ as defined in \eqref{eq:Z_eta}, one finds that
\beq
\label{eq:A_1}
 A_1(\et{F}) = \dot Z^{(1)}(\et{F})
\eeq
with the dot denoting the derivative \wrt $\eta$.

Still assuming that $\Xi S_\ia = 0$, the Weyl anomaly $A(\et{S_\ia})$ can be determined from \eqref{eq:WeylAnomaly} order by order in the interaction as\footnote{$A(1) = 0$, by the vanishing of the anomaly of a c-number functional and the field independence of the anomaly.}
\begin{align}
\label{eq:A_S_int}
 A( S_\ia ) & = \cT^{-1} ( \Xi \cT(S_\ia) ) \,, \\
\label{eq:A_S_int_S_int}
 A( S_\ia \otimes S_\ia ) & = \ibar \cT^{-1} \left( \Xi \cT( S_\ia \otimes S_\ia ) - 2 \cT( A( S_\ia) \otimes S_\ia ) \right) \,,
\end{align}
and similarly for higher orders. Note that the arguments of $\cT^{-1}$ are always local functionals, so that $\cT^{-1}$ is well-defined and yields a local functional. 

Let us now relate the general Weyl anomaly $A(\et{F})$ to the trace anomaly. On the classical level, the stress tensor associated to an action $S$ is defined by
\beq
\label{eq:StressTensor}
 \delta S = - \frac{1}{2} \int T_{\mu \nu} \delta g^{\mu \nu} \vol_g = \frac{1}{2} \int T^{\mu \nu} \delta g_{\mu \nu} \vol_g
\eeq
for a variation of the action \wrt the metric. By considering the variations of $S_0$ and $S_\ia$ separately, one obtains the decomposition $T_{\mu \nu} = T^0_{\mu \nu} + T^\ia_{\mu \nu}$ of the stress tensor. We will particularly be interested in the trace 
\beq
\label{eq:T_f}
 T(f) = \int f g^{\mu \nu} T_{\mu \nu} \vol_g
\eeq
of the stress tensor, smeared with a test function $f$, which can also be obtained by choosing $\delta g_{\mu \nu} = 2 f g_{\mu \nu}$ in \eqref{eq:StressTensor}. 
According to Bogoliubov's formula \eqref{eq:BogoliubovsFormula}, the corresponding observable in the interacting theory is $\cT^\ia( T(f) ) = \cR(T(f) ; \eti{S_\ia})$.
Classically, i.e., at $\cO(\hbar^0)$, this vanishes on-shell, so an on-shell non-vanishing $\cT^\ia( T(f) )$ constitutes a trace anomaly.
We compute
\begin{align}
 \cT^\ia( T(f)) & = \cR(T^0(f); \eti{S_\ia}) + \cR(T^\ia(f); \eti{S_\ia}) \nn \\
 & = \cT(\eti{S_\ia})^{-1} \star \cR( \eti{S_\ia}; T^0(f) ) \nn \\
\label{eq:R_T_f}
 & \qquad + \cT(\eti{S_\ia})^{-1} \star \cT(T^0(f)) \star \cT(\eti{S_\ia}) + \cR(T^\ia(f); \eti{S_\ia}) \,.
\end{align}
Here we used the general identity
\beq
 \cR(F; \eti{G}) = \cT(\eti{G})^{-1} \star \cR(\eti{G}; F) + \cT(\eti{G})^{-1} \star \cT(F) \star \cT(\eti{G}) \,,
\eeq
which is a direct consequence of \eqref{eq:RetardedProduct}.
Now the trace anomaly $\cT(T^0(f))$ of the free theory is on-shell a c-number, so it commutes on-shell \wrt the $\star$ product. This can be used to simplify the second term on the right hand side of \eqref{eq:R_T_f}. To treat the first term, we use perturbative agreement \eqref{eq:TO_PA} with $h_{\mu \nu} = 2 f g_{\mu \nu}$ (so that $\delta_h S_0 = T^0(f)$), so that we obtain (recall that $\simeq$ denotes equality in the on-shell algebra)
\begin{align}
 \cT^\ia( T(f) ) & \simeq \cT(T^0(f)) - i \hbar \cT( \eti{S_\ia} )^{-1} \star \delta^\ret_h \cT( \eti{S_\ia} ) \nn \\
 & \qquad - \cT(\eti{S_\ia})^{-1} \star \cT( T^\ia(f) \otimes \eti{S_\ia} ) + \cR(T^\ia(f); \eti{S_\ia}) \nn \\
 & \simeq \cT(T^0(f)) - i \hbar \cT( \eti{S_\ia} )^{-1} \star \delta^\ret_h \cT( \eti{S_\ia} ) \,.
\end{align}
The first term on the \rhs is the contribution from the free theory, whereas the second term is the contribution due to the interaction. Let us examine this term more closely. We already argued that $\delta^\ret_h$ for $h_{\mu \nu} = 2 f g_{\mu \nu}$ coincides on-shell with $\delta^\weyl_f$, \cf \eqref{eq:gamma_Omega_simeq_tau_ret} and the discussion below \eqref{eq:delta_s_f}. It follows that we can write
\beq
\label{eq:T_int_T_f}
 \cT^\ia( T(f) ) \simeq \cT(T^0(f)) + \cT^\ia( A_f( \et{S_\ia}) ) \,,
\eeq
with $A_f$ the anomaly as introduced in \eqref{eq:A_f} (corresponding to $A$ with the ghost field $\xi$ replaced by $f$).

In some cases (in particular in the $\phi^4$ theory studied below), $A_f(\et{S_\ia})$ contains a term of the form of the trace of the stress tensor $T(f)$ (as defined in \eqref{eq:T_f}), i.e., we can write
\beq
 A_f(\et{S_\ia}) = \gamma T(f) + \widetilde{A_f(\et{S_\ia})} \,,
\eeq
with $\gamma$ at least of $\cO(\hbar)$ and $\widetilde{A_f(\et{S_\ia})}$ depending on $\phi$ and (covariant derivatives of) curvature tensors, but not on derivatives of $\phi$.
Then we can rewrite \eqref{eq:T_int_T_f} as
\beq
\label{eq:T_int_T_f_with_gamma}
 \cT^\ia( T(f) ) \simeq \frac{1}{1- \gamma} \left( \cT(T^0(f)) + \cT^\ia( \widetilde{A_f( \et{S_\ia})} ) \right) \,.
\eeq

Let us now consider the behaviour of the interacting trace anomaly under renormalization group transformations in the sense of \cite{HollandsWaldRG}. We consider time-ordered products $\cT^{(\eta)}$ which are obtained as in \eqref{eq:T_eta} by a scaling transformation from a given prescription $\cT$ for time-ordered products (so in particular $\cT^{(1)} = \cT$). Denoting by $\cT^{(\eta) \ia}$ and $A^{(\eta)}$ the corresponding interacting time-ordered product and anomaly, we have
\beq
 \cT^{(\eta) \ia}( T(f) ) \simeq \cT^{(\eta)} (T^0(f)) + \cT^{(\eta) \ia}( A^{(\eta)}_f( \et{S_\ia} ) ) \,.
\eeq
Now the anomaly $A^{(\eta)}$ \wrt $\cT^{(\eta)}$ can be determined in terms of the original anomaly $A$ and the redefinition map $Z^{(\eta)}$ according to \eqref{eq:Xi_Z}. Computing the derivative of $A^{(\eta)}(\et{S_\ia})$ \wrt $\eta$ at $\eta = 1$, we obtain
\begin{align}
 \dot A^{(1)}_{f}(\et{S_\ia}) & = \delta^\weyl_f \dot Z^{(1)}(\et{S_\ia}) + A_f( \dot Z^{(1)}(\et{S_\ia}) \otimes \et{S_\ia}) - \dot Z^{(1)}( A_f(\et{S_\ia}) \otimes \et{S_\ia}) \nn \\
 & = \delta^\weyl_f A_1( \et{S_\ia} ) + A_f( A_1( \et{S_\ia} ) \otimes \et{S_\ia} ) - A_1( A_f( \et{S_\ia}) \otimes \et{S_\ia} ) \nn \\
 & = \delta^\weyl_1 A_f( \et{S_\ia} ) \,.
\end{align}
Here we used \eqref{eq:A_1} and the consistency condition \eqref{eq:ConsistencyCondition}, which, when expressed in terms of bosonic variations $f_1$, $f_2$ reads
\beq
 \delta^\weyl_{f_1} A_{f_2}(\et{S_\ia}) - \delta^\weyl_{f_2} A_{f_1}(\et{S_\ia}) = - A_{f_1}( A_{f_2}(\et{S_\ia}) \otimes \et{S_\ia}) + A_{f_2}( A_{f_1}(\et{S_\ia}) \otimes \et{S_\ia}) \,.
\eeq
Assuming that \eqref{eq:A_f_HomogeneousScaling} indeed holds (\cf the discussion preceding that equation), i.e., the anomaly $A_f( \et{S_\ia} )$ scales homogeneously under constant scale transformation, we obtain
\beq
 \dot A^{(1)}_{f}(\et{S_\ia}) = 0 \,,
\eeq
so that in this sense the trace anomaly is invariant under the renormalization group flow.

We now turn to the consideration of the behavior of arbitrary interacting observables under local scale transformations. 
Still assuming that $\delta^\weyl_f S_\ia = 0$, we obtain, using the definition \eqref{eq:BogoliubovsFormula} of interacting fields, the definition \eqref{eq:A_f} of the Weyl anomaly, and the fact that $\delta^\weyl_f$ fulfills the Leibniz rule \wrt the $\star$ product,
\beq
\label{eq:Xi_T_int_exp_F}
 \delta^\weyl_f \cT^\ia( F ) = \cT^\ia( \delta^\weyl_f F + A_f(F \otimes \et{S_\ia}) ) + \ibar \cT^\ia( F \otimes A_f(\et{S_\ia}) ) - \ibar \cT^\ia( A_f(\et{S_\ia}) ) \star \cT^\ia(F) \,.
\eeq
When the last two terms on the \rhs cancel for all $F$, then we can read this equation as stating that a local scale transformation on an interacting observable $\cT^\ia(F)$ corresponding to a local functional $F$ is implemented by the local action $F \mapsto \delta^\weyl_f F + A_f( F \otimes \et{S_\ia})$ on $F$. 
Here ``local'' can be understood in both of the following two senses: 
i) $\delta^\weyl_f \cT^{\ia}(F)$ vanishes if the supports of $f$ and $F$ do not overlap. ii) $\delta^\weyl_f \cT^\ia(F)$ commutes with $\cT^\ia(G)$ whenever the supports of $F$ and $G$ are spacelike related. 
Hence, when the last two terms on the \rhs of \eqref{eq:Xi_T_int_exp_F} cancel for all $F$, one would call the interacting theory conformal, and $A_f( F \otimes \et{S_\ia} )$ would be interpreted as the anomalous scaling of the observable $F$. 

Now the last two terms on the \rhs of \eqref{eq:Xi_T_int_exp_F} cancel for all $F$ if and only if the interacting contribution $A_f(\et{S_\ia})$ to the trace anomaly is a c-number. However, requiring that $A_f(\et{S_\ia})$ must be a c-number in order for the interacting theory to be conformal seems overly restrictive. After all, it might be that a proper definition of $\delta^\weyl_f$ on interacting observables needs to take quantum corrections into account, which could cancel the
last two terms on the \rhs of \eqref{eq:Xi_T_int_exp_F}. One possibility for this is a term $\gamma T(f)$ in $A_f(\et{S_\ia})$, with $T(f)$ as defined in \eqref{eq:T_f} and $\gamma$ a constant at least of $\cO(\hbar)$. Such a term does indeed typically appear in $A_f(\et{S_\ia})$ (above, we already discussed the effect of such a term on the trace anomaly). Now as a consequence of perturbative agreement, we have, for an arbitrary variation $h_{\mu \nu}$ of the metric, \cite{BgInd}
\beq
 \delta^\ret_h \cT^\ia( \eti{F} ) = \ibar \cT^\ia( \delta_h F \otimes \eti{F} ) + \ibar \cT^\ia( \eti{F} \otimes \delta_h S ) - \ibar \cT^\ia( \delta_h S ) \star \cT^\ia(\eti{F}) \,.
\eeq
Hence, in case that $A(\et{S_\ia})$ equals $\gamma T(f)$ up to c-number terms, by choosing the infinitesimal metric variation to be conformal, i.e., $h_{\mu \nu} = 2 f g_{\mu \nu}$, one can write \eqref{eq:Xi_T_int_exp_F} as
\beq
 \left( \delta^\weyl_f - \gamma \delta^\ret_{2 f g} \right) \cT^\ia( F ) = \cT^\ia( \delta^\weyl_f F - \gamma \delta_{2 f g} F + A_f( F \otimes \et{S_\ia}) ) \,.
\eeq
Recalling that $\delta^\weyl_f$ and $\delta^\ret_{2 f g}$ coincide on the on-shell algebra, we can even write
\beq
 \delta^\weyl_f \cT^\ia( F ) \simeq \cT^\ia( \delta^{\weyl,\hbar}_f F ) \, ,
\eeq
where the \emph{quantum Weyl transformation}
\beq
\delta^{\weyl,\hbar}_f F \coloneq (1-\gamma)^{-1} \{ \delta^\weyl_f F - \gamma \delta_{2 f g} F + A_f( F \otimes \et{S_\ia}) \} \,,
\eeq
differs from $\delta^\weyl_f$ by terms which are at least of order $\hbar$. Hence, an interacting contribution $A_f(\et{S_\ia})$ to the trace anomaly of the form $\gamma T(f)$ plus c-number terms seems to be a sensible characterization of conformal field theories in the present framework. This also seems to coincide with the definition adopted in \cite{Fortin:2012hn} in the path integral framework.

An example of such a term $\gamma T(f)$ that is known in the literature appears for the massless Sine--Gordon model in two spacetime dimensions \cite{BabujianKarowski, AlbertiSchlesierZahn, FrobCadamuroSineGordon}. The classical stress tensor in Minkowski spacetime is given by
\begin{equation}
\label{eq:sinegordon_stresstensor}
T_{\mu\nu} = \del_\mu \phi \del_\nu \phi - \frac{1}{2} \eta_{\mu\nu} \eta^{\rho\sigma} \del_\rho \phi \del_\sigma \phi + 2 g \eta_{\mu\nu} \cos(\beta \phi) \,,
\end{equation}
and even though the interaction $S_\ia = 2 g \cos(\beta \phi)$ is not conformally invariant, the free massless scalar field is such that our general framework is applicable (with $\Xi S_\ia$ non-vanishing). That the trace of the stress tensor \eqref{eq:sinegordon_stresstensor} receives an anomalous contribution even in flat space has been discovered in the form factor program \cite{BabujianKarowski}, and was then verified first to one loop \cite{AlbertiSchlesierZahn} and afterwards non-perturbatively \cite{FrobCadamuroSineGordon} in the algebraic approach. It turned out that $\gamma = - \tfrac{\hbar \beta^2}{8 \pi}$ is one-loop exact, such that the full trace reads
\begin{equation}
\cT^\ia( T(f) ) = 4 \left( 1 - \frac{\hbar \beta^2}{8 \pi} \right) g \int f(x) \cT^\ia ( \cos[\beta \phi(x)] ) \vol(x) \,.
\end{equation}

\section{Cohomological analysis of \texorpdfstring{$\phi^4$}{φ⁴} theory}
\label{eq:phi4_Cohomology}

As an example, we consider conformally coupled $\phi^4$ theory, i.e., the action
\beq
\label{eq:Action}
 S = \int \left( - \frac{1}{2} \nabla_\mu \phi \nabla^\mu \phi - \frac{1}{12} R \phi^2 - \frac{\lambda}{4!} \phi^4 \right) \vol \,,
\eeq
leading to the equation of motion
\beq
\label{eq:eom}
 \Box \phi - \frac{1}{6} R \phi = \frac{\lambda}{6} \phi^3 \,,
\eeq
where we recall that $\Box = \nabla^\mu \nabla_\mu$. The corresponding classical stress tensor is
\beq
 T_{\mu \nu} = \nabla_\mu \phi \nabla_\nu \phi - \frac{1}{2} g_{\mu \nu} \nabla^\lambda \phi \nabla_\lambda \phi - \frac{1}{6} \nabla_\mu \nabla_\nu \phi^2 + \frac{1}{6} g_{\mu \nu} \nabla^\lambda \nabla_\lambda \phi^2 + \frac{1}{6} G_{\mu \nu} \phi^2 - \frac{\lambda}{4!} g_{\mu \nu} \phi^4 \,.
\eeq
Classically, it is on-shell both conserved and traceless. We denote by $T_{\mu \nu}^0$ the free part (quadratic in the fields), and by $T_{\mu \nu}^\ia$ the interacting part (of higher order in the fields). For the free part $T^0_{\mu \nu}$, we have
\begin{align}
 \nabla^\mu T^0_{\mu \nu} & = \nabla_\nu \phi \left( \Box - \frac{1}{6} R \right) \phi \,, &
 g^{\mu \nu} T^0_{\mu \nu} & = \phi \left( \Box - \frac{1}{6} R \right) \phi \,,
\end{align}
so it is classically both conserved and on-shell \wrt the free equations of motion, i.e., with the non-linear term on the \rhs of \eqref{eq:eom} removed. For later purposes, it is convenient to introduce the symbol
\beq
\label{eq:lambdabar}
 \lambdabar \defeq - \frac{\lambda}{4!} \,,
\eeq
so that the interaction Lagrangian is given by $L_\ia = \lambdabar \phi^4$.

Let us first review the treatment of the free part of the trace anomaly, i.e., $\cT( T^0)$. It occurs because the Hadamard parametrix $H$, which is used to define Wick powers according to \eqref{eq:HadamardPointSplit}, is a bi-solution of the Klein-Gordon equation only modulo smooth remainder terms. Even worse, in a Hadamard point split prescription, also conservation of the free stress tensor does not hold, as one finds that
\beq
 \cT ( \nabla^\mu T_{\mu \nu}^0 ) \simeq \nabla_\nu Q \,,
\eeq
where $Q$ is a local curvature functional \cite{MorettiStressEnergy}. Choosing $\phi^2$ and $\phi \nabla_\mu \nabla_\nu \phi$ as a basis of ``Leibniz independent'' Wick squares involving up to two derivatives (recall the discussion above \eqref{eq:nabla_T_phi_2}), one can achieve, by a redefinition of $\cT( \phi \nabla_\mu \nabla_\nu \phi )$, a conserved free stress tensor \cite{HollandsWaldStress}. We emphasize that $\cT(\phi^2)$ does not need to be redefined for that purpose. 

Having achieved an on-shell conserved free stress tensor (in the free theory), we can also (possibly by redefining time-ordered products involving a free stress tensor as one of the factors) achieve perturbative agreement \eqref{eq:TO_PA}, and thus conservation of the interacting stress tensor (in the interacting theory) \cite{HollandsWaldStress}. Furthermore, after the above redefinition, one finds a trace of $\cT(T_{\mu \nu}^0)$ which is on-shell of the form 
\beq
\label{eq:FreeTraceAnomaly}
 \cT(T^0) \simeq - a_0 \cE_4 + c_0 C^2 + b_0 \Box R \,,
\eeq
where \cite{FrobZahn2019} (in agreement with the classical result \cite{Christensen76})
\begin{align}
 a_0 & = \frac{\hbar}{5760 \pi^2} \,, &
 c_0 & = \frac{\hbar}{1920 \pi^2} \,,
\end{align}
and $\cE_4$, $C^2$ are the Euler density and the square of the Weyl tensor, respectively:
\begin{align}
 C^2 & = R_{\mu \nu \lambda \rho} R^{\mu \nu \lambda \rho} - 2 R_{\mu \nu} R^{\mu \nu} + \frac{1}{3} R^2 \,, \\
 \cE_4 & = R_{\mu \nu \lambda \rho} R^{\mu \nu \lambda \rho} - 4 R_{\mu \nu} R^{\mu \nu} + R^2 \,.
\end{align}
The term $\Box R$ can be removed by a further redefinition of $\cT( \phi \nabla_\mu \nabla_\nu \phi )$ which does not affect the on-shell conservation of the free stress tensor \cite{HollandsWaldStress, FrobZahn2019} (again, $\cT(\phi^2)$ need not be redefined). 
Concretely, the redefinition necessary to turn the Hadamard point-split definition of time-ordered products into one in which the free stress tensor is conserved and $b_0$ in \eqref{eq:FreeTraceAnomaly} vanishes, is given by
\begin{multline}
\label{eq:Z_phi_nabla_nabla_phi}
 Z( \phi \nabla_\mu \nabla_\nu \phi ) = \frac{\hbar}{2880 \pi^2} g_{\mu \nu} \left( R^{\alpha \beta \gamma \delta} R_{\alpha \beta \gamma \delta} - R^{\alpha \beta} R_{\alpha \beta} + \Box R \right) \\ + \frac{\hbar}{8640 \pi^2} \left( \nabla_\mu \nabla_\nu R + \frac{1}{2} g_{\mu \nu} \Box R - R R_{\mu \nu} + \frac{1}{4} g_{\mu \nu} R^2 \right) \,.
\end{multline}

Let us at this point comment on a different approach to treating the free stress tensor suggested in \cite{MorettiStressEnergy}. There, the classical expression for the stress tensor is modified by adding a term proportional to the equations of motion (which vanishes classically) in order to guarantee the conservation of its quantum counterpart. While the results obtained for the free trace anomaly coincide, the approach pursued here (achieving a conserved quantum stress tensor by modifying time-ordered products) has the advantage that it guarantees, by the results of \cite{HollandsWaldStress}, that perturbative agreement can be fulfilled and that, as a consequence, also the interacting stress tensor is conserved. The approach of \cite{MorettiStressEnergy} was recently further investigated in \cite{CosteriDappiaggiGoi}, where it was shown that in a modified scheme taking the interaction into account, conservation of the interacting stress tensor can be obtained up to second order in the interaction, but only on spacetimes for which the coinciding point limit of a certain Hadamard coefficient is constant (such as maximally symmetric spacetimes).

We now turn to the discussion of the interacting contribution $A(\et{S_\ia})$ to the trace anomaly. According to the general discussion above, it is useful for that purpose to determine the behaviour under $\Xi$ of all possible functionals of mass dimension four and ghost numbers $0$ and $1$. With \cite[App.~D]{WaldGR}
\begin{subequations}
\label{eqs:Xi_R_identities}
\begin{align}
 \Xi R_{\mu \nu \lambda}{}^\rho & = 2 \delta^\rho_{[\mu} \nabla_{\nu]} \nabla_\lambda \xi - 2 g_{\lambda [\mu} \nabla_{\nu]} \nabla^\rho \xi \,, \\
\label{eq:Xi_R_mu_nu}
 \Xi R_{\mu \nu} & = - 2 \nabla_\mu \nabla_\nu \xi - g_{\mu \nu} \Box \xi \,, \\
 \Xi R & = - 2 \xi R - 6 \Box \xi \,, \\
 \label{eq:Xi_Box}
 \Xi \Box & = - 2 \xi \Box + 2 \nabla_\mu \xi \nabla^\mu \,,
\end{align}
\end{subequations}
and $\Xi \vol = 4 \xi \vol$, one finds that (here and in the following we use the convention that the differential operators $\nabla_\mu$, $\Box$ only act on the directly following factor, so that, for example, $\Box \xi \phi^2 = \phi^2 \Box \xi$)
\begin{subequations}
\label{eqs:L_i}
\begin{align}
 L_1 & = R_{\mu \nu \lambda \rho} R^{\mu \nu \lambda \rho} \vol & \Xi L_1 & = - 8 R^{\mu \nu} \nabla_\mu \nabla_\nu \xi \vol \,, \\
 L_2 & = R_{\mu \nu} R^{\mu \nu} \vol & \Xi L_2 & = - (4 R^{\mu \nu} + 2 g^{\mu \nu} R) \nabla_\mu \nabla_\nu \xi \vol \,, \\
 L_3 & = R^2 \vol & \Xi L_3 & = - 12 \Box \xi R \vol \,, \\
 L_4 & = \Box R \vol & \Xi L_4 & = \left( - 2 \nabla_\mu ( \nabla^\mu \xi R) - 6 \Box \Box \xi \right) \vol \,, \\
 L_5 & = \phi \Box \phi \vol & \Xi L_5 & = - \Box \xi \phi^2 \vol \,, \\
 L_6 & = R \phi^2 \vol & \Xi L_6 & = - 6 \Box \xi \phi^2 \vol \,, \\
 L_7 & = \phi^4 \vol & \Xi L_7 & = 0 \,, \\
 L_8 & = \Box \phi^2 \vol & \Xi L_8 & = - 2 \nabla_\mu ( \nabla^\mu \xi \phi^2) \vol \,.
\end{align}
\end{subequations}
Multiplying the above densities with an adiabatic cut-off function $\chi$, integrating, then applying $\Xi$ and finally setting $\chi = 1$ on the support of $\xi$ then yields\footnote{Regarding analogous results in \cite{BonoraEtAl83}, we note that apparently a different sign convention for $\Box$ was used there and that the result given there for $\Xi \int \chi \Box R \vol$ is incorrect.}
\begin{subequations}
\label{eqs:Xi_L_0}
\begin{align}
\label{eq:Xi_L_0_R}
 \Xi \int \chi R_{\mu \nu \lambda \rho} R^{\mu \nu \lambda \rho} \vol = \Xi \int \chi R_{\mu \nu} R^{\mu \nu} \vol = \frac{1}{3} \Xi \int \chi R^2 \vol & = - 4 \int \Box \xi R \vol \,, \\
\label{eq:Xi_L_0_phi}
 \Xi \int \chi \phi \Box \phi \vol = \frac{1}{6} \Xi \int \chi R \phi^2 \vol & = - \int \Box \xi \phi^2 \vol \,, \\
 \Xi \int \chi \Box R \vol = \Xi \int \chi \Box \phi^2 \vol = \Xi \int \chi \phi^4 \vol & = 0 \,.
\end{align}
\end{subequations}
Analogously, taking the fermionic nature of $\xi$ into account,
\begin{subequations}
\label{eqs:Xi_L_1}
\begin{align}
\label{eq:Xi_L_1_R}
 \Xi \int \xi R_{\mu \nu \lambda \rho} R^{\mu \nu \lambda \rho} \vol = \Xi \int \xi R_{\mu \nu} R^{\mu \nu} \vol = \frac{1}{3} \Xi \int \xi R^2 \vol & = 4 \int \xi \Box \xi R \vol \,, \\
 \Xi \int \xi \phi \Box \phi \vol = \frac{1}{6} \Xi \int \xi R \phi^2 \vol & = \int \xi \Box \xi \phi^2 \vol \,, \\
 \Xi \int \xi \Box R \vol = \Xi \int \xi \Box \phi^2 \vol = \Xi \int \xi \phi^4 \vol & = 0 \,.
\end{align}
\end{subequations}
One can easily see that the two functionals $\int \xi \Box \xi R \vol$ and $\int \xi \Box \xi \phi^2 \vol$ span the space of functionals of ghost number two and mass dimension four.

From the consistency condition \eqref{eq:ConsistencyCondition} and \eqref{eqs:Xi_L_1}, it thus follows that at lowest non-vanishing order in $\hbar$ (denoted by $m$ here), the interacting trace anomaly is of the form\footnote{From \eqref{eq:Xi_L_1_R} it follows that $\int \xi (- a \cE_4 +c C^2 ) \vol$ is the most general linear combination of curvature squares multiplied with $\xi$ which is annihilated by $\Xi$.}
\beq
\label{eq:A_m_exp_S_int}
 A^{(m)}(\et{S_\ia}) = \int \xi \left( - a^{(m)} \cE_4 + c^{(m)} C^2 + \gamma^{(m)} T + \beta^{(m)} \phi^4 + d^{(m)} \Box R + \alpha^{(m)} \Box \phi^2 \right) \vol \,,
\eeq
where
\beq
 T = \phi \left( \Box - \frac{1}{6} R \right) \phi - \frac{\lambda}{6} \phi^4
\eeq
vanishes on-shell and coincides with the trace of the stress tensor. Furthermore, as seen in \eqref{eqs:Xi_L_0}, the last two terms in \eqref{eq:A_m_exp_S_int} are $\Xi$ exact, so one can perform a redefinition of time-ordered products as discussed around \eqref{eq:Z_exp_S_int} above in order to set $d^{(m)} = \alpha^{(m)} = 0$, i.e., we arrive at
\beq
\label{eq:A_m_exp_S_int_wo_trivial}
 A^{(m)}(\et{S_\ia}) = \int \xi \left( - a^{(m)} \cE_4 + c^{(m)} C^2 + \gamma^{(m)} T + \beta^{(m)} \phi^4 \right) \vol \,.
\eeq
We will perform this redefinition explicitly in Section~\ref{sec:phi4_Computation} and in particular show how to do it in such a way that perturbative agreement is preserved.

A natural question is now whether this can be extended to higher order in $\hbar$, i.e., whether one can achieve \eqref{eq:A_m_exp_S_int_wo_trivial} to any order in $\hbar$. Let us thus assume that we have achieved
\beq
\label{eq:A_k_exp_S_int_wo_trivial}
 A^{(k)}(\et{S_\ia}) = \int \xi \left( - a^{(k)} \cE_4 + c^{(k)} C^2 + \gamma^{(k)} T + \beta^{(k)} \phi^4 \right) \vol
\eeq
for all $k \leq m$, and consider the anomaly $A^{(m+1)}(\et{S_\ia})$ at the next order in $\hbar$. Applying the consistency condition \eqref{eq:ConsistencyCondition} and the inductive assumption, we arrive at
\beq
\label{eq:Xi_A_m+1}
 \Xi A^{(m+1)}(\et{S_\ia}) = - A( \{ \tilde \gamma^{(m)} T(\xi) + \lambdabar^{-1} \tilde \beta^{(m)} S_\ia(\xi) \} \otimes \et{S_\ia})\big\rvert_{\cO(\hbar^{m+1})} \,,
\eeq
with the notation defined in \eqref{eq:T_f}, \eqref{eq:S_int_chi}, and \eqref{eq:lambdabar}.
Here we used that the anomaly with a c-number factor vanishes and introduced $\tilde \gamma^{(m)}$ as the sum of the anomaly coefficients $\gamma^{(k)}$ up to (and including) $\cO(\hbar^m)$ (and analogously for $\tilde \beta^{(m)}$).

Let us deal with the $\tilde \gamma^{(m)}$ term first. As a consequence of perturbative agreement, the anomaly fulfills the same relation \eqref{eq:delta_ret_h_Z} as the redefinition map $Z$, for any variation $h_{\mu \nu}$ of the background metric. It follows that\footnote{\label{ft:A_xi_F}Note the sign change in the first term on the \rhs which is due to the need to pull the (now fermionic) $h_{\mu \nu}$ into the anomaly from the left: We have, for bosonic $F$, $A(\xi(x) F(x) \otimes \et{H}) = - \xi (x) A( F(x) \otimes \et{H})$, which follows from Footnote~\ref{ft:T_xi_Phi} and $\xi$ and $\Xi$ being fermionic.}
\beq
\label{eq:A_T_xi_exp_S_int}
 A^{(k)}( T(\xi) \otimes \et{S_\ia} ) = - \delta_{2 \xi g} A^{(k)}( \et{S_\ia} ) + A^{(k)}( T^0(\xi) ) \,.
\eeq
Regarding the last term on the r.h.s., we have
\beq
\label{eq:Xi_T_T_0_xi}
 \Xi \cT( T^0(\xi) ) = \cT( A( T^0(\xi) ) ) = A( T^0(\xi) ) \,,
\eeq
where in the first step we used the results in \eqref{eqs:L_i} and in the second step the fact that the anomaly of a Wick square is a c-number (so that the time-ordered product acts trivially). Due to the latter fact, we only need to consider the c-number part of the \lhs of \eqref{eq:Xi_T_T_0_xi}, which is nothing but $\Xi$ applied to the trace anomaly of the free theory smeared with $\xi$, which vanishes by \eqref{eq:Xi_L_1_R}. In the first term on the \rhs of \eqref{eq:A_T_xi_exp_S_int}, we need to consider $k \leq m$, as $\tilde \gamma^{(m)}$ is at least of $\cO(\hbar)$. Using the scaling behaviour of $\Box$ and $\xi$, the fermionic nature of $\xi$, and the inductive assumption, one easily checks that also the first term on the \rhs of \eqref{eq:A_T_xi_exp_S_int} vanishes.\footnote{Note that the field $\phi$ transforms trivially under $\delta_h$, so that the fermionic nature of $\xi$ is necessary to achieve a vanishing variation of the $T$ and $\phi^4$ terms.}

Hence, it suffices to consider the $\tilde \beta^{(m)}$ term in \eqref{eq:Xi_A_m+1}, i.e., the \rhs of \eqref{eq:Xi_A_m+1} vanishes if
\beq
\label{eq:A_k_chi_phi_4}
 A^{(k)}( S_\ia(\xi) \otimes \et{S_\ia} ) = 0
\eeq
for $k \leq m$.
If \eqref{eq:A_k_chi_phi_4} holds for $k \leq m$, then the anomaly at $\cO(\hbar^{m+1})$ is $\Xi$ closed, so in particular again of the form \eqref{eq:A_m_exp_S_int}. By an appropriate redefinition of time-ordered products, one may then also achieve $d^{(m+1)} = \alpha^{(m+1)} = 0$. We have thus proven that if \eqref{eq:A_k_exp_S_int_wo_trivial} and \eqref{eq:A_k_chi_phi_4} hold for $k \leq m$, then there is a renormalization scheme such that \eqref{eq:A_k_exp_S_int_wo_trivial} also holds for $k \leq m+1$.

This result has the obvious shortcoming that in order to proceed to the next order in $\hbar$, we need that also \eqref{eq:A_k_chi_phi_4} holds for $k \leq m+1$. From a superficial analysis, one might conclude that \eqref{eq:A_k_exp_S_int_wo_trivial} actually implies \eqref{eq:A_k_chi_phi_4}: By the absence in \eqref{eq:A_k_exp_S_int_wo_trivial} of terms that can be written as multiple of $\Box \xi$ (due to the removal of the cohomologically trivial $d$ and $\alpha$ term), it seems as if smearing one of the interaction terms with $\xi$, one can only arrive at $\xi^2$ terms for $A^{(m+1)}( S_\ia(\xi) \otimes \et{S_\ia})$, which vanish. However, we have to keep in mind how \eqref{eq:A_k_exp_S_int_wo_trivial} is to be understood. In the interaction terms on the \lhs an adiabatic cutoff function $\chi$ is introduced, which is then set equal to $1$ on the support of $\xi$ (which is assumed to be compact). As long as one keeps the cut-off function general, there may be supplementary terms involving derivatives of $\chi$. For example, a contribution ($c$ being a numerical coefficient)
\beq
\label{eq:A_f_phi_4}
 c \tilde \chi \left( \Box \xi R - \xi \Box R \right) \vol
\eeq
to the anomaly\footnote{We have here introduced a bosonic cutoff function $\tilde \chi$ distinct from the cutoff $\chi$ implicitly contained in $S_\ia$. The support of $\tilde \chi$ is contained in the region where $\chi$ is equal to $1$, so that $\chi$ can be disregarded.} $A^{(m+1)}( S_\ia(\tilde \chi) \otimes S_\ia^{\otimes l} )$ would be a total derivative for $\tilde \chi = 1$ on the support of $\xi$ (and thus not contribute to $A^{(m+1)}( S_\ia^{\otimes (l+1)} )$), while under the replacement $\tilde \chi \to \xi$ it gives rise to the anomaly (the sign change occurring for the reason discussed in Footnote~\ref{ft:A_xi_F})
\beq
\label{eq:A_chi_phi_4_example}
 A^{(m+1)}( S_\ia(\xi) \otimes S_\ia^{\otimes l} ) = - c \xi \Box \xi R \vol \,.
\eeq
Hence, \eqref{eq:A_k_exp_S_int_wo_trivial} does not straightforwardly imply \eqref{eq:A_k_chi_phi_4}. Even worse, assume that \eqref{eq:A_k_exp_S_int_wo_trivial} holds for $k \leq m+1$, but \eqref{eq:A_k_chi_phi_4} only holds for $k \leq m$ and the anomaly at the next order in $\hbar$ is given by \eqref{eq:A_chi_phi_4_example}. 
Then we cannot remove the anomaly \eqref{eq:A_chi_phi_4_example} without introducing an anomaly $A^{(m+1)}( S_\ia^{\otimes (l+1)} )$. 
To see this, we note that in order to remove the anomaly \eqref{eq:A_chi_phi_4_example} to this order in $\hbar$, we have to perform the redefinition (\cf \eqref{eq:Xi_Z}, \eqref{eq:Xi_L_1_R} and the Lagrangians defined in \eqref{eqs:L_i})
\beq
 Z( S_\ia(\xi) \otimes S_\ia^{\otimes l} ) = \int \xi \left( d_1 L_1 + d_2 L_2 + d_3 L_3 \right) \,,
\eeq
where $d_1$, $d_2$, $d_3$ are chosen such that $\frac{1}{4} d_1 + \frac{1}{4} d_2 + \frac{1}{12} d_3 = - c$. However, by field independence, we then also have to redefine
\beq
 Z( L_\ia \otimes S_\ia^{\otimes l} ) = d_1 L_1 + d_2 L_2 + d_3 L_3 \,,
\eeq
which results in
\beq
 A^{(k)}(S_\ia^{\otimes (l+1)}) = c \int \Box \xi R \vol \,,
\eeq
i.e., we have reintroduced the unwanted $\Box R$ term. The analogous result holds if $R$ on the \rhs of \eqref{eq:A_chi_phi_4_example} is replaced by $\phi^2$. We thus see that, as already pointed out in \cite{Osborn1991}, terms in the anomaly involving derivatives of the coupling ``constant'' need to be taken into account in a complete cohomological analysis.

We do not see an argument ruling out ``total derivatives'' of the form \eqref{eq:A_f_phi_4} in the anomaly, and results in the literature \cite{HathrellScalar, Osborn1991} suggest that they are indeed present. However, using the explicit result that $A(\phi^4) = 0$ can be achieved (see next section), we can show that the anomaly can be brought to the form given on the \rhs of \eqref{eq:A_m_exp_S_int_wo_trivial} up to third order in the interaction (to any order in $\hbar$). For this argument, it is useful to modify the recursive scheme, such that the recursion in $\hbar$ is performed at a fixed order in the interaction. As at a fixed order in the interaction only a finite power of $\hbar$ can occur (due to \eqref{eq:Z_F} which also holds for $A$), this recursion is finished after a finite number of steps, and one proceeds to the next order in the interaction.

Proceeding like this for $A(S_\ia)$, i.e., removing the $b$ and $\alpha$ terms recursively in $\hbar$ (see next section for the concrete calculation), one arrives at $A(S_\ia(\xi)) = 0$. In fact, we will see that we can even achieve the stronger
\beq
\label{eq:A_S_int_tilde_chi}
 A(S_\ia(\tilde \chi)) = 0 \,,
\eeq
which we assume from now on.
However, by the consistency condition \eqref{eq:ConsistencyCondition}, already $A(S_\ia) = 0$ suffices to conclude
\beq
 \Xi A( S_\ia \otimes S_\ia ) = - A ( A( S_\ia \otimes S_\ia) ) \,.
\eeq
Hence, proceeding inductively in $\hbar$, we can remove the $b$ and $\alpha$ terms and arrive at $A(S_\ia \otimes S_\ia)$ of the form given on the \rhs of \eqref{eq:A_m_exp_S_int_wo_trivial}. This is performed explicitly in the next section. In order to be able to deal with $A(S_\ia \otimes S_\ia \otimes S_\ia)$ in the following step, we now have to argue that
\beq
\label{eq:A_S_int_xi_S_int}
 A(S_\ia(\xi) \otimes S_\ia) = 0 \,.
\eeq
For this, we first consider the most general form of $A( S_\ia(\tilde \chi) \otimes S_\ia)$ (with the cutoff $\chi$ implicit in the second factor $S_\ia$ to be equal to one on the support of $\tilde \chi$), consistent with $A( S_\ia \otimes S_\ia)$ of the form given on the \rhs of \eqref{eq:A_m_exp_S_int_wo_trivial}, namely
\begin{align}
\label{eq:A_S_int_f_S_int}
 & \lambdabar^{-2} A( S_\ia(\tilde \chi) \otimes S_\ia) = \int \xi \tilde \chi \left( - a_2 \cE_4 + c_2 C^2 + \gamma_2 T^0 + \beta_2 \phi^4 \right) \vol \\ 
 & \quad + \int \tilde \chi \left[ d_1 ( \Box \xi \phi^2 - \xi \Box \phi^2 ) + d_2 \Box ( \xi \phi^2 ) + d_3 ( \Box \xi R - \xi \Box R ) + d_4 \Box ( \xi R ) + d_5 \nabla_\mu ( \nabla_\nu \xi R^{\mu \nu} ) \right] \vol\,. \nn
\end{align}
The first term on the \rhs corresponds to the \rhs of \eqref{eq:A_m_exp_S_int_wo_trivial}, except that we replaced the trace $T$ of the stress tensor by its free part $T^0$ and included $\tilde \chi$ in the integral. We also used the expansion
\beq
\label{eq:a_Expansion}
 a = \sum_{k = 1}^\infty \frac{\lambdabar^k}{k!} a_k \,,
\eeq
of the anomaly coefficients in terms of the coupling constant (analogously for $c$, $\beta$, $\gamma$). 
The expression in square brackets in the second term on the \rhs of \eqref{eq:A_S_int_f_S_int} is the most general total derivative that is compatible with power counting. Hence, the expression above is the most general one compatible with the requirement that $A(S_\ia(\tilde \chi) \otimes S_\ia )$ equals $A(S_\ia \otimes S_\ia )$ when $\tilde \chi = 1$ on the support of $\xi$. By replacing $\tilde \chi$ with $\xi$, we also see that with the above ansatz (note again the sign change explained in Footnote~\ref{ft:A_xi_F})
\beq
\label{eq:A_S_int_xi_S_int_ds}
 A( S_\ia(\xi) \otimes S_\ia ) = \int \xi \Box \xi \left( (d_2 - d_1) \phi^2 + (d_4 - d_3) R \right) \vol \,.
\eeq

We now want to derive constraints on the coefficients $d_i$ from the consistency condition. 
Using \eqref{eq:A_S_int_f_S_int} and the results in \eqref{eqs:Xi_R_identities}, \eqref{eqs:L_i},
\begin{multline}
\label{eq:Xi_A_S_int_f_S_int}
 \Xi A( S_\ia(\tilde \chi) \otimes S_\ia ) = \lambdabar^2 \int \tilde \chi \big[ 8 a_2 \xi G^{\mu \nu} \nabla_\mu \nabla_\nu \xi + 2 (d_2 - d_1) \xi \nabla_\mu ( \nabla^\mu \xi \phi^2 ) \\ 
 + (d_4 - d_3) ( 2 \xi \nabla_\mu ( \nabla^\mu \xi R ) + 6 \xi \Box \Box \xi ) + (3 d_5 + 12 d_4) \nabla_\mu \xi \nabla^\mu \Box \xi \big] \vol \,.
\end{multline}
By \eqref{eq:A_S_int_tilde_chi} and the consistency condition \eqref{eq:ConsistencyCondition}, this must be equal to
\beq
 - A (A (S_\ia(\tilde \chi) \otimes S_\ia )) \,.
\eeq
As \eqref{eq:A_S_int_tilde_chi} implies, by field independence, also $A( \phi^2 ) = 0$, the only possibly contributing term is the $\gamma$ term, i.e.,
\beq
 - \lambdabar^{-2} A (A (S_\ia(\tilde \chi) \otimes S_\ia )) = - \gamma_2 A( T^0( \tilde \chi \xi ) ) = a_0 \gamma_2 \Xi \int \cE_4 \tilde \chi \xi \vol = - 8 a_0 \gamma_2 \int \tilde \chi \xi G^{\mu \nu} \nabla_\mu \nabla_\nu \xi \vol \,.
\eeq
Here we used \eqref{eq:Xi_T_T_0_xi}, which also holds with $\xi$ replaced by $\tilde \chi \xi$. Comparison with \eqref{eq:Xi_A_S_int_f_S_int} (and noting that the terms in there are linearly independent), we conclude that the consistency condition implies that $a_2 = - a_0 \gamma_2$, $d_2 = d_1$, $d_4 = d_3$, and $d_5 = - 4 d_4$. Hence, comparing with \eqref{eq:A_S_int_xi_S_int_ds}, we have established \eqref{eq:A_S_int_xi_S_int}, but also found an interesting constraint among $a$ and $\gamma$. In particular, from \eqref{eq:T_int_T_f_with_gamma}, it follows that the effective $a$ coefficient in the interacting theory is given by
\beq
 \tilde a \defeq \frac{1}{1 - \gamma} \left( \sum_{k = 0}^\infty \frac{\lambdabar^k}{k!} a_k \right) = a_0 + \frac{\lambdabar^2}{2} \left( \gamma_2 a_0 + a_2 \right) + \cO(\lambdabar^3) = a_0 + \cO(\lambdabar^3) \,,
\eeq
where in the last step we used the relation found above. Thus, the vanishing of $\cO(\lambda^2)$ corrections to $\tilde a$ follows from \eqref{eq:A_S_int_tilde_chi} and the consistency condition.\footnote{The same conclusion can also be derived in the framework of \cite{Osborn1991}.}

We can now proceed to third order in the interaction. By the consistency condition \eqref{eq:ConsistencyCondition} and $A(S_\ia) = 0$,
\beq
 \Xi A(S_\ia \otimes S_\ia \otimes S_\ia) = - A( A( S_\ia \otimes S_\ia \otimes S_\ia) ) - 3 A( A(S_\ia \otimes S_\ia) \otimes S_\ia) \,.
\eeq
By the same arguments that we used in the discussion above, the second term on the \rhs is a linear combination of $A( S_\ia(\xi) \otimes S_\ia)$ and $A( S_\ia(\xi))$,\footnote{In the above discussion (starting at \eqref{eq:A_T_xi_exp_S_int}), we argued that $A( T(\xi) \otimes \et{S_\ia})$ vanishes at the appropriate order in $\hbar$. However, when expanding in the coupling constant, one should instead consider $A(T^0(\xi) \otimes S_\ia)$. Using the same arguments as above, one can express it in terms of $A( S_\ia(\xi) )$.} which both vanish. It follows that, proceeding order by order in $\hbar$, the third order anomaly $A(S_\ia \otimes S_\ia \otimes S_\ia)$ can be brought to the form given on the \rhs of \eqref{eq:A_m_exp_S_int_wo_trivial}. In particular, there is then no $\Box \phi^2$ term at this order. As discussed in the Introduction, this is to be contrasted with the results of \cite{BrownCollins, HathrellScalar}, where a vanishing $\Box \phi^2$ term at this order is only possible by invoking non-perturbative effects.

\section{Computing the Weyl anomaly in \texorpdfstring{$\phi^4$}{φ⁴} theory}
\label{sec:phi4_Computation}

We now turn to explicitly computing the trace anomaly in $\phi^4$ theory up to second order in the interaction. We begin by computing the anomaly $A(S_\ia)$ to first order in the interaction, and find a non-vanishing $\alpha$ (in the notation used in \eqref{eq:A_m_exp_S_int}). Instead of directly removing this trivial anomaly, we continue with the second order anomaly $A(S_\ia \otimes S_\ia)$, and find that it contains terms such as $R \phi^2$ and $R^2$. We then see that upon removal of the first order anomaly $A(S_\ia)$, these conformally non-invariant terms indeed drop out.

In order to compute the anomaly $A( S_\ia )$, we have to consider the usual Hadamard point-split prescription for the Wick power, i.e., according to \eqref{eq:HadamardPointSplit},
\beq
\label{eq:T_phi_4}
 \cT(\phi^4(x)) = \phi^4(x) + 6 \hbar \phi^2(x) (w - H)(x, x) + 3 \hbar^2 (w - H)^2(x, x) \,.
\eeq
According to \eqref{eq:W_Omega}, the two-point function $w$ scales homogeneously,
\beq
\label{eq:Xi_W}
 \Xi w(x, x') = - (\xi(x) + \xi(x')) w(x, x') \,.
\eeq
This is not the case for the Hadamard parametrix, which can lead to an anomaly $A(S_\ia)$. Concretely, the Hadamard parametrix can be locally (for nearby $x$, $x'$) expressed as
\beq
\label{eq:H}
 H(x, x') = \frac{1}{8 \pi^2} \left( \frac{u}{\sigma_\eps} + v \log \frac{\sigma_\eps}{\Lambda^2} \right)
\eeq
where $\sigma_\eps$ is Synge's \emph{world function} (defined below) equipped with the $\rmi \eps$ prescription
\beq
 \sigma_\eps(x, x') = \sigma(x, x') + \rmi (t(x) - t(x')) \eps 
\eeq
for some time function $t$ on $M$, $\Lambda$ is an arbitrarily chosen scale and $u(x, x')$ and $v(x, x')$ are smooth functions. More precisely, $u(x, x')$ is the square root of the van Vleck--Morette determinant $\Delta$ and $v(x,x')$ can be written as a series $v = \sum_k \sigma^k v_k$ in terms of Hadamard coefficients $v_k(x, x')$ and the world function $\sigma(x, x')$, which is defined as
\beq
\label{eq:sigma}
 \sigma(x, x') = \frac{1}{2} \int_0^1 g_{\mu \nu}(z(\tau)) \frac{\ud z^\mu}{\ud \tau}\frac{\ud z^\mu}{\ud \tau} \ud \tau \,,
\eeq
with $\tau \mapsto z(\tau)$ the geodesic such that $z(0) = x$ and $z(1) = x'$ (or vice versa). Both for the van Vleck--Morette determinant $\Delta$ (and thus also $u$) and for $v$, there are useful expansions near coinciding points in terms of $\sigma^\mu = \nabla^\mu \sigma$ (which coincides with $(x-x')^\mu$ in Cartesian coordinates on Minkowski space) \cite{PoissonPoundVega, DecaniniFolacci08}:
\begin{align}
\label{eq:Delta}
\Delta & = 1 + \frac{1}{6} R_{\mu\nu} \sigma^\mu \sigma^\nu - \frac{1}{12} \nabla_{(\rho} R_{\mu\nu)} \sigma^\mu \sigma^\nu \sigma^\rho \nn \\
&\quad+ \left( \frac{1}{72} R_{(\mu\nu} R_{\rho\sigma)} + \frac{1}{40} \nabla_{(\rho} \nabla_\sigma R_{\mu\nu)} + \frac{1}{180} \tensor{R}{_{(\rho|\alpha|\sigma|\beta|}} \tensor{R}{_\mu^\alpha_{\nu)}^\beta} \right) \sigma^\mu \sigma^\nu \sigma^\rho \sigma^\sigma + \cO( (\sigma^\mu)^5) \,, \\
\label{eq:v}
 v & = \frac{1}{720} \left( - 2 \tensor{R}{_\mu^{\alpha\beta\gamma}} \tensor{R}{_{\nu\alpha\beta\gamma}} - 2 \tensor{R}{^{\alpha\beta}} \tensor{R}{_{\mu\alpha\nu\beta}} + 4 \tensor{R}{_\mu^\alpha} \tensor{R}{_{\nu\alpha}} - 3 \Box R_{\mu\nu} + \nabla_\mu \nabla_\nu R \right) \sigma^\mu \sigma^\nu \nn \\
&\quad+ \frac{1}{720} \left( \tensor{R}{^{\mu\nu\rho\sigma}} \tensor{R}{_{\mu\nu\rho\sigma}} - R^{\mu\nu} R_{\mu\nu} + \Box R \right) \sigma + \cO( (\sigma^\mu)^3 ) \,.
\end{align}
In these expressions, all curvature tensors are evaluated at $x$.

As we already know the action of $\Xi$ on curvature tensors, it remains to determine its action on $\sigma$, for which one finds, as described in Appendix~\ref{app:Proofs},
\begin{align}
\label{eq:Xi_sigma}
 \Xi \sigma & = 2 \sigma \sum_{k = 0}^\infty \frac{(-1)^k}{(k+1)!} (\sigma \cdot \nabla)^k \xi(x) + \cO( (\sigma^\mu)^\infty ) \\
 & = [ \xi(x) + \xi(x') ] \sigma - \frac{1}{6} \sigma ( \sigma \cdot \nabla )^2 \xi(x) + \frac{1}{12} \sigma ( \sigma \cdot \nabla )^3 \xi(x) - \frac{1}{40} \sigma ( \sigma \cdot \nabla )^4 \xi(x) + \cO((\sigma^\mu)^7) \,, \nn
\end{align}
where
\beq
\label{eq:sigma_cdot_nabla}
 (\sigma \cdot \nabla)^k \defeq \sigma^{\mu_1} \dots \sigma^{\mu_k} \nabla_{\mu_1} \dots \nabla_{\mu_k} \,.
\eeq
From the fact that $\sigma^\mu = g^{\mu \nu} \del_\nu \sigma$ and $\Xi$ commutes with $\del_\mu$, we obtain
\begin{align}
 \Xi \sigma^\mu & = \sigma \left[ 1 - \frac{1}{3} \sigma \cdot \nabla + \frac{1}{12} ( \sigma \cdot \nabla )^2 - \frac{1}{60} ( \sigma \cdot \nabla )^3 \right] \nabla^\mu \xi(x) \\
&\quad- \sigma^\mu \left[ \sigma \cdot \nabla - \frac{1}{3} ( \sigma \cdot \nabla )^2 + \frac{1}{12} ( \sigma \cdot \nabla )^3 - \frac{1}{60} ( \sigma \cdot \nabla )^4 \right] \xi(x) \nn \\
&\quad+ \frac{1}{12} \tensor{R}{^\mu_{\rho\nu\sigma}} \sigma \sigma^\rho \sigma^\sigma \left[ 1 - \frac{7}{15} \sigma \cdot \nabla \right] \nabla^\nu \xi(x) - \frac{1}{30} \nabla_\alpha \tensor{R}{^\mu_{\rho\nu\sigma}} \sigma \sigma^\alpha \sigma^\rho \sigma^\sigma \nabla^\nu \xi(x) + \cO( (\sigma^\mu)^6 ) \,, \nn
\end{align}
where also
\beq
 \nabla_\mu \sigma_\nu = g_{\mu\nu} - \frac{1}{3} R_{\mu\rho\nu\sigma} \sigma^\rho \sigma^\sigma + \frac{1}{12} \nabla_\alpha R_{\mu\rho\nu\sigma} \sigma^\rho \sigma^\sigma \sigma^\alpha + \cO( (\sigma^\mu)^4 )
\eeq
was used (which extends a result given in \cite{PoissonPoundVega}). It follows that
\begin{align}
 \Xi \Delta & = - \frac{1}{3} \left( \sigma^\mu \sigma^\nu + g^{\mu\nu} \sigma \right) \nabla_\mu \nabla_\nu \xi + \frac{1}{6} \sigma^\rho \left( \sigma^\mu \sigma^\nu + g^{\mu\nu} \sigma \right) \nabla_\rho \nabla_\mu \nabla_\nu \xi \nn \\
&\quad- \frac{1}{20} \sigma^\rho \sigma^\sigma \left( \sigma^\mu \sigma^\nu + g^{\mu\nu} \sigma \right) \nabla_\rho \nabla_\sigma \nabla_\mu \nabla_\nu \xi \nn \\
&\quad- \frac{1}{90} \left( \sigma \sigma^\mu \sigma^\rho \tensor{R}{^\nu_\rho} + 5 \sigma^\mu \sigma^\nu \sigma^\rho \sigma^\sigma R_{\rho\sigma} + 5 \sigma \sigma^\rho \sigma^\sigma g^{\mu\nu} R_{\rho\sigma} + 2 \sigma \sigma^\rho \sigma^\sigma \tensor{R}{_\rho^\mu_\sigma^\nu} \right) \nabla_\mu \nabla_\nu \xi \nn \\
&\quad+ \frac{1}{30} \sigma \sigma^\mu \sigma^\nu \left( \nabla_\mu \tensor{R}{_\nu^\rho} - \nabla^\rho R_{\mu\nu} \right) \nabla_\rho \xi + \cO ( (\sigma^\mu)^5 ) \,, \\
 \Xi v & = - \left[ \xi(x) + \xi(x') \right] v + \cO( (\sigma^\mu)^3 ) \,,
\end{align}
which implies that
\begin{align}
\label{eq:Xi_H}
 \Xi H(x, x') & = - [ \xi(x) + \xi(x') ] H(x,x') - \frac{1}{48 \pi^2} \left( 1 - \frac{1}{2} \sigma^\mu \nabla_\mu + \frac{3}{20} \sigma^\mu \sigma^\nu \nabla_\mu \nabla_\nu \right) \Box \xi \nn \\
&\quad+ \frac{1}{4 \pi^2} v(x,x') \xi(x) + \frac{1}{480 \pi^2} \sigma^\mu \sigma^\nu \left( \nabla_\mu \tensor{R}{_\nu^\rho} - \nabla^\rho R_{\mu\nu} \right) \nabla_\rho \xi \nn \\
&\quad- \frac{1}{2880 \pi^2} \left( 2 \sigma^\mu \sigma^\rho \tensor{R}{^\nu_\rho} + 5 \sigma^\rho \sigma^\sigma g^{\mu\nu} R_{\rho\sigma} + 4 \sigma^\rho \sigma^\sigma \tensor{R}{_\rho^\mu_\sigma^\nu} \right) \nabla_\mu \nabla_\nu \xi + \cO ( (\sigma^\mu)^3 ) \,.
\end{align}
We have here determined $\Xi H$ including the second order in $\sigma^\mu$. For the determination of $A(S_\ia)$ which we will do in the following, already terms of $\cO(\sigma^\mu)$ can be dismissed, as we will be concerned in the limit of coinciding points. However, the expansion to higher orders will become relevant for the determination of $A(S_\ia \otimes S_\ia)$ below.

We now turn to the evaluation of $A(S_\ia)$. Using \eqref{eq:T_phi_4}, \eqref{eq:Xi_W}, and \eqref{eq:Xi_H} (and $\Xi \vol = 4 \xi \vol$), we obtain
\beq
\label{eq:Xi_T_phi_4}
 \Xi \cT( \phi^4(x) \vol(x) ) = \frac{\hbar}{8 \pi^2} \Box \xi \left[ \phi^2(x) + \hbar ( w - H )(x, x) \right] \vol(x) = \frac{\hbar}{8 \pi^2} \Box \xi \cT(\phi^2(x) \vol(x)) \,,
\eeq 
such that, using \eqref{eq:A_S_int},
\beq
\label{eq:A_phi_4}
 A( \phi^4 ) = \frac{\hbar}{8 \pi^2} \Box \xi \phi^2 \,.
\eeq
In complete analogy, one also finds
\begin{align}
\label{eq:Xi_T_phi_3}
 \Xi \cT( \phi^3(x) \vol(x) ) & = \xi(x) \cT( \phi^3(x) \vol(x) ) + \frac{\hbar}{16 \pi^2} \Box \xi(x) \cT( \phi(x) \vol(x)) \,, \\
\label{eq:Xi_T_phi_2}
 \Xi \cT( \phi^2(x) \vol(x) ) & = 2 \xi(x) \cT( \phi^2(x) \vol(x) ) + \frac{\hbar}{48 \pi^2} \Box \xi(x) \vol(x) \,,
\end{align}
so in particular
\beq
\label{eq:A_phi_2}
 A( \phi^2 ) = \frac{\hbar}{48 \pi^2} \Box \xi \,,
\eeq
a result that will be relevant below. From \eqref{eq:A_phi_4}, we see that an $\alpha$ term (in the notation used in \eqref{eq:A_m_exp_S_int}) is present at first order in the interaction. Before we remove it, we first consider the anomaly in the next order of the interaction.

We now turn to the evaluation of the anomaly $A(S_\ia \otimes S_\ia)$ to second order in the interaction. We do this in the Hadamard point split scheme,
but bear in mind that we still need to perform redefinitions in order to achieve a conserved stress tensor and remove the trivial anomaly at first order in the interaction that we just found.

According to \eqref{eq:A_S_int_S_int}, in order to compute $A(S_\ia \otimes S_\ia)$, we need to consider $\Xi \cT(S_\ia \otimes S_\ia)$. With \eqref{eq:TO_Formal}, we formally have
\begin{align}
\label{eq:T_phi_4_phi_4}
 \cT( \phi^4(x) \otimes \phi^4(x') ) & = \cT( \phi^4(x) ) \cT( \phi^4(x') ) + 16 \hbar \cT( \phi^3(x) ) \cT (\phi^3(x') ) w_\feyn(x, x') \nn \\
 & \quad + 72 \hbar^2 \cT( \phi^2(x) ) \cT( \phi^2(x') ) w_\feyn(x, x')^2 \nn \\
 & \quad + 96 \hbar^3 \cT( \phi(x) ) \cT( \phi(x') ) w_\feyn(x, x')^3 + 24 \hbar^4 w_\feyn(x, x')^4 \,.
\end{align}
This is only formal as the Feynman propagator is a distribution, so taking products is in general not possible. It turns out that powers of the Feynman propagator are indeed well-defined, but only up to coinciding points, i.e., as distributions on $M^2 \setminus \{ (x, x) \mid x \in M \}$ \cite{BrunettiFredenhagenScalingDegree}. The basic idea for extending these distributions to all of $M^2$ is as follows: One splits the Feynman propagator as
\beq
 w_\feyn(x, x') = H_\feyn(x, x') + W(x, x')
\eeq
into the \emph{Feynman parametrix} $H_\feyn$ and a smooth remainder $W$. The Feynman parametrix is defined as the Hadamard parametrix $H(x, x')$, \cf \eqref{eq:H}, but with a different $\rmi \eps$ prescription, namely by the replacement $\sigma_\eps \to \sigma + \rmi \eps$. The remainder $W(x, x')$ is in fact the same as occurring in the Hadamard point-split scheme, $W = w - H$. We will see below how, given the specific form of the Feynman parametrix $H_\feyn$, one can define its powers (extend them to distributions defined in a neighborhood of coinciding points). Thus, also the powers of $w_\feyn$ are then defined by ($W$ and its powers are smooth, so that the multiplication in this expression is well-defined)
\beq
\label{eq:W_F_k}
 w_\feyn^k = \sum_{l = 0}^k \binom{k}{l} H_\feyn^l W^{k-l} \,.
\eeq

What will be relevant for our consideration is the inhomogeneous scaling of the resulting distributions: As $w(x, x')$, also the Feynman propagator $w_\feyn$ scales homogeneously, i.e., as in \eqref{eq:Xi_W}. It follows that the same is true for its powers $w_\feyn(x,x')^2$ where they are defined, i.e., for $x \neq x'$, we have
\beq
 \Xi w_\feyn(x, x')^k = - k [\xi(x) + \xi(x')] w_\feyn(x, x')^k \,.
\eeq
However, in the extension process, homogeneous scaling is in general violated. By the above analysis, the violation terms must be supported at coinciding points, i.e., they must be (derivatives of) $\delta$ distributions. With \eqref{eq:W_F_k}, we obtain
\beq
 \Xi w_\feyn^k = \sum_{l = 0}^k \binom{k}{l} \left( W^{k-l} \Xi H_\feyn^l + (k-l) H_\feyn^l W^{k-l-1} \Xi W \right) .
\eeq
By the homogeneous scaling of $w_\feyn$, we have
\beq
 \Xi W = - [ \xi(x) + \xi(x') ] ( H_\feyn + W ) - \Xi H_\feyn \,,
\eeq
and using this in the above, we arrive at
\beq
\label{eq:Xi_W_F_k_loc}
 \Xi w_\feyn^k = - k [ \xi(x) + \xi(x') ] w_\feyn^k + \sum_{l = 2}^k \binom{k}{l} W^{k-l} \Xi_\loc H_\feyn^l \,,
\eeq
where we used the definition
\beq
\label{eq:Xi_loc}
 \Xi_\loc H_\feyn^l \defeq \Xi H_\feyn^l + l [\xi(x) + \xi(x') ] H_\feyn^l - l H_\feyn^{l-1} ( \Xi H_\feyn + [ \xi(x) + \xi(x') ] H_\feyn ) \,.
\eeq
In \eqref{eq:Xi_loc}, the expression in brackets in the last term is smooth, so the product in the last term is well-defined. $\Xi_\loc H_\feyn^l$ extracts the local contributions to inhomogeneous scaling, i.e., $\delta$ distributions and derivatives thereof.

We can now relate $\Xi_\loc H_\feyn^l$ to the anomaly at second order in the interaction. Using \eqref{eq:Xi_T_phi_4}, \eqref{eq:Xi_T_phi_3}, \eqref{eq:Xi_T_phi_2} to recombine the inhomogeneously scaling terms of the Wick power $\cT( \phi^k )$, we obtain, by applying $\Xi$ to both sides of \eqref{eq:T_phi_4_phi_4},
\begin{align}
 & \Xi \cT ( \phi^4 \vol(x) \otimes \phi^4 \vol(x') ) \\
 & = \frac{\hbar}{8 \pi^2} \Box \xi(x) \cT( \phi^2 \vol(x) \otimes \phi^4 \vol(x')) + \frac{\hbar}{8 \pi^2} \Box \xi(x') \cT( \phi^4 \vol(x) \otimes \phi^2 \vol(x')) \nn \\
 & \quad + 72 \hbar^2 \, \cT( \phi^2 \vol(x) ) \cT( \phi^2 \vol(x') ) \left( \Xi w_\feyn^2 + 2 [ \xi(x) + \xi(x') ] w_\feyn^2 \right) \nn \\
 & \quad + 96 \hbar^3 \, \cT( \phi \vol(x) ) \cT( \phi \vol(x') ) \left( \Xi w_\feyn^3 + 3 [ \xi(x) + \xi(x') ] w_\feyn^3 \right) \nn \\
 & \quad + 24 \hbar^4 \vol(x) \vol(x') \left( \Xi w_\feyn^4 + 3 [ \xi(x) + \xi(x') ] w_\feyn^4 \right) \,. \nn
\end{align}
Using \eqref{eq:Xi_W_F_k_loc}, this can be expressed as 
\begin{align}
\label{eq:A_S_int_S_int_Step2}
 & \Xi \cT ( \phi^4 \vol(x) \otimes \phi^4 \vol(x') ) \\
 & = \frac{\hbar}{8 \pi^2} \Box \xi(x) \cT( \phi^2 \vol(x) \otimes \phi^4 \vol(x')) + \frac{\hbar}{8 \pi^2} \Box \xi(x') \cT( \phi^4 \vol(x) \otimes \phi^2 \vol(x')) \nn \\
 & \quad + 72 \hbar^2 \, \cT( \phi^2 \vol(x) ) \cT( \phi^2 \vol(x') ) \Xi_\loc H_\feyn^2 \nn \\
 & \quad + 96 \hbar^3 \, \cT( \phi \vol(x) ) \cT( \phi \vol(x') ) \left( \Xi_\loc H_\feyn^3 + 3 W \Xi_\loc H_\feyn^2 \right) \nn \\
 & \quad + 24 \hbar^4 \vol(x) \vol(x') \left( \Xi_\loc H_\feyn^4 + 4 W \Xi_\loc H_\feyn^3 + 6 W^2 \Xi_\loc H_\feyn^2 \right) \,. \nn
\end{align}
With \eqref{eq:A_S_int_S_int} and \eqref{eq:A_phi_4}, we thus obtain, suppressing the variables $x$, $x'$ in the integrals on the right hand side,
\begin{align}
\label{eq:A_S_int_S_int_Step3}
 \lambdabar^{-2} A ( S_\ia \otimes S_\ia ) & = 72 \rmi \hbar \, \cT^{-1} \left( \int_{M^2} \left[ \cT( \phi^2 ) \cT( \phi^2 ) + 4 \cT( \phi ) \cT( \phi ) W + 2 W^2 \right] \Xi_\loc H_\feyn^2 \vol \vol \right) \nn \\
 & \quad + 96 \rmi \hbar^2 \, \cT^{-1} \left( \int_{M^2} \left[ \cT( \phi ) \cT( \phi ) + W \right] \Xi_\loc H_\feyn^3 \vol \vol \right) \nn \\
 & \quad + 24 \rmi \hbar^3 \int_{M^2} \Xi_\loc H_\feyn^4 \vol \vol \,.
\end{align}
From the relation of the degree of singularity of a distributions and the ambiguity of its extension discussed below, it follows that we can express $\Xi_\loc H_\feyn^2$ and $\Xi_\loc H_\feyn^3$ as
\begin{align}
\label{eq:Xi_loc_H_F_2_General}
 \Xi_\loc H_\feyn(x,x')^2 & = A \xi(x) \delta(x, x') \,, \\
 \Xi_\loc H_\feyn(x,x')^3 & = \left( B_0 R(x) \xi(x) + B_1 \Box \xi(x) \right) \delta(x, x') + C \nabla^\mu \xi(x) \nabla_\mu \delta(x, x') + D \Box \delta(x, x') \,,
\end{align}
with numerical coefficients $A, B_0, B_1, C$ and $D$.
Furthermore, as in \eqref{eq:A_S_int_S_int_Step3} we integrate $\Xi_\loc H_\feyn^4$ over $M^2$ without any further (cutoff) function, we only need $\Xi_\loc H_\feyn^4$ up to total derivatives. We can thus write it as
\beq
 \Xi_\loc H_\feyn(x,x')^4 = E(x) \xi(x) \delta(x, x') \,, 
\eeq
with $E(x)$ a linear combination of curvature squares and $\Box R$. Recalling that $W = w - H$ is the difference also occurring in the definition of Wick powers, we see that the expressions in square brackets in \eqref{eq:A_S_int_S_int_Step3} indeed combine into Wick powers, such that the action of $\cT^{-1}$ yields
\begin{align}
\label{eq:A_S_int_S_int_Step4}
 \lambdabar^{-2} A ( S_\ia \otimes S_\ia ) & = 72 \rmi \hbar A \int_M \xi \phi^4 \vol \\
 & \quad + 96 \rmi \hbar^2 \int_M \xi \left( D \phi \Box \phi + B_0 R \phi^2 + \left( B_1 - \frac{C}{2} \right) \Box \phi^2 \right) \vol + 24 \rmi \hbar^3 \int_M \xi E \vol \,. \nn 
\end{align}

To determine the coefficients $A$, $B_0$, $B_1$, $C$, $D$, $E(x)$, we need to consider the inhomogeneous scaling of the extension of powers of $H_\feyn$. In order to motivate the subsequent definition of powers of $H_\feyn$ (i.e., extension to a distribution defined in a neighborhood of coinciding points), we first consider, as an elementary example, the distribution $x_+^{-1} \defeq \theta(x) x^{-1}$ on $\R \setminus \{ 0 \}$ (with $\theta$ the step function). One easily checks that on a test function $f(x)$ which vanishes in a neighborhood of $x = 0$, one can also express it as
\beq
 x_{+, \Lambda}^{-1}(f) = - \int_0^\infty f'(x) \ln \frac{x}{\Lambda} \ud x
\eeq
with $\Lambda > 0$ an arbitrary scale. The point is that the logarithm is integrable near $0$, so the above is also well-defined for arbitrary test functions $f$ (not necessarily vanishing in a neighborhood of $0$). However, for such a general test function, the choice of $\Lambda$ \emph{does} matter: We have
\beq
 x_{+, \Lambda'}^{-1}(f) - x_{+, \Lambda}^{-1}(f) = \ln \frac{\Lambda}{\Lambda'} f(0),
\eeq
where the \rhs corresponds to a Dirac $\delta$ (evaluated on the test function $f$).
We can now also define the even more divergent $x_+^{-k}$ for $k \geq 2$ as a distribution on $\R$ by repeatedly differentiating $x_{+, \Lambda}^{-1}$. The ambiguity related to the choice of $\Lambda$ is then related to derivatives of the Dirac $\delta$ distribution. We thus see that by expressing an inverse as the derivative of a logarithm (and higher inverse powers as derivatives thereof), we can extend the domain of definition of a distribution. As the logarithm does not scale homogeneously, homogeneous scaling is violated in the extension. Furthermore, there are ambiguities in the process (related to a change of the scale $\Lambda$), which amount to (derivatives of) Dirac $\delta$ distributions. More generally \cite{Steinmann, BahnsWrochna}, for a distribution $u$ on $\R^n \setminus \{ 0 \}$ with a degree of divergence (in the above example of $x_+^{-k}$, the degree of divergence would be $k$) smaller than $n$, there is a unique extension $\tilde u$ to a distribution on $\R^n$ with the same degree of divergence. If the degree of divergence of $u$ is finite but greater or equal to $n$, then extensions $\tilde u$ preserving the degree of divergence exist, but are not unique. The ambiguity consist in (derivatives of) $\delta$ distributions, with the number of derivatives bounded by the degree of divergence of $u$ minus $n$.\footnote{This has natural generalization to distributions on $X \setminus Y$ with $X$ a manifold and $Y$ a submanifold thereof \cite{BrunettiFredenhagenScalingDegree, DangExtension}. As we will be dealing with distributions on $M^2 \setminus \{ (x, x) \mid x \in M \}$, this would be the natural mathematical framework to use. However, by fixing $x'$ and expressing $x$ in normal coordinates around $x'$, we can convert the relevant distributions to distributions defined on (open subsets of) $\R^4 \setminus \{ 0 \}$.}

Let us now turn to the most divergent term in $H_\feyn(x,x')^2$, namely, $\frac{1}{(\sigma + \rmi \eps)^2}$. Using the relations \cite{PoissonPoundVega}
\begin{align}
 \sigma_\mu \sigma^\mu & = 2 \sigma \,, &
 \sigma^\mu \nabla_\mu \Delta & = ( 4 - \nabla_\mu \sigma^\mu ) \Delta \,,
\end{align}
one easily checks that one can rewrite it (for $x \neq x'$) as\footnote{An analogous trick was used in \cite{HollandsYM} in terms of normal coordinates for $x$ around $x'$, and in \cite{FreedmanJohnsonLatorre} in flat space.}
\begin{equation}
\label{eq:sigma-2} 
 \frac{1}{( \sigma + \rmi \eps )^2} = - \frac{1}{2} \cD\left( \frac{ \ln \frac{ \sigma + \rmi \eps }{\Lambda^2 } }{ \sigma + \rmi \eps } \right) \,,
\end{equation}
where
\beq
 \cD = \Box + \nabla^\mu \ln \Delta \nabla_\mu \,.
\eeq
The important point is that the distribution on which $\cD$ acts in \eqref{eq:sigma-2} has a degree of divergence of two for $x \to x'$, so that (according to the above) one can unambiguously extend it. Once this is done, the action of $\cD$ yields another distribution defined on a neighborhood of $x = x'$. Having thus defined $\frac{1}{(\sigma + \rmi \eps)^2}$, we can define higher powers by
\begin{align}
 \frac{1}{( \sigma + \rmi \eps )^3} & = \frac{1}{4} \cD \frac{1}{ (\sigma + \rmi \eps)^2 } \,, &
 \frac{1}{( \sigma + \rmi \eps )^4} & = \frac{1}{12} \cD \frac{1}{ (\sigma + \rmi \eps)^3 } \,.
\end{align}

From the explicit form \eqref{eq:Delta}, \eqref{eq:v} of $u = \Delta^{\frac{1}{2}}$ and $v$ appearing in the form \eqref{eq:H} of the Feynman parametrix (recall that $H_\feyn$ is of the same form as $H$, but with a different $\rmi \eps$ prescription), it follows that
\beq
 H_\feyn^2 = - \frac{1}{128 \pi^4} \Box\left( \frac{ \ln \frac{ \sigma + \rmi \eps }{\Lambda^2 } }{ \sigma + \rmi \eps } \right) + \cO( ( \sigma^\mu )^{-2} \ln \sigma ) \,,
\eeq
where the neglected terms cannot contribute to the inhomogeneous scaling as their degree of divergence is smaller than four. 
With \eqref{eq:Xi_Box}, \eqref{eq:Xi_sigma}, we obtain
\begin{align}
 \Xi H_\feyn^2(x,x') & = - \frac{1}{64 \pi^4} \xi(x) \Box\left( \frac{ \ln \frac{ \sigma + \rmi \eps }{\Lambda^2 } }{ \sigma + \rmi \eps } \right) - \frac{1}{128 \pi^4} \Box \left[ \frac{\Xi \sigma}{\sigma} \frac{1 - \ln \frac{ \sigma + \rmi \eps }{\Lambda^2 } }{\sigma + \rmi \eps} \right] + \cO( ( \sigma^\mu )^{-3} \ln \sigma ) \\
 & = - 2 \xi(x) H_\feyn^2(x, x') - \frac{1}{128 \pi^4} \Box \left[ [ \xi(x) + \xi(x') ] \frac{1 - \ln \frac{ \sigma + \rmi \eps }{\Lambda^2 } }{\sigma + \rmi \eps} \right] + \cO( ( \sigma^\mu )^{-3} \ln \sigma ) \nn \\
 & = - 2 [ \xi(x) + \xi(x') ] H_\feyn^2(x, x') - \frac{1}{128 \pi^4} [ \xi(x) + \xi(x') ] \Box \frac{1}{\sigma + \rmi \eps} + \cO( ( \sigma^\mu )^{-3} \ln \sigma ) \,. \nn
\end{align}
With
\beq
 \Box \frac{1}{\sigma + \rmi \eps} = 8 \pi^2 \left( \Box - \frac{1}{6} R \right) H_\feyn + \cO((\sigma^\mu)^{-2}) = 8 \pi^2 \rmi \delta(x, x') + \cO((\sigma^\mu)^{-2}) \,,
\eeq
we thus obtain
\beq
 \Xi_\loc H_\feyn^2(x, x') = - \frac{\rmi}{8 \pi^2} \xi(x) \delta(x, x') \,,
\eeq
i.e., in the notation introduced in \eqref{eq:Xi_loc_H_F_2_General},
\beq
 A = - \frac{\rmi}{8 \pi^2} \,.
\eeq
For the higher powers of $H_\feyn$ one proceeds similarly, i.e., commuting $\Xi$ through $\cD$, for which we conveniently use computer algebra~\cite{xact} to derive
\begin{align}
 B_0 & = - \frac{\rmi}{1536 \pi^4} \,, &
 B_1 & = - \frac{11 \rmi}{1536 \pi^4} \,, &
 C & = - \frac{\rmi}{256 \pi^4} \,, &
 D & = - \frac{\rmi}{256 \pi^4}
\end{align}
and
\beq
 E(x) = - \frac{\rmi}{1105920 \pi^6} \left( 3 R_{\mu \nu \lambda \rho} R^{\mu \nu \lambda \rho} + 12 R_{\mu \nu} R^{\mu \nu} + 5 R^2 + 58 \Box R \right) \,.
\eeq
We thus arrive at
\begin{align}
 A( S_\ia \otimes S_\ia ) & = \lambdabar^2 \int_M \xi \biggl[ \frac{9}{\pi^2} \hbar \phi^4 + \frac{3}{8 \pi^4} \hbar^2 \phi \left( \Box + \frac{1}{6} R \right) \phi + \frac{1}{2 \pi^2} \hbar^2 \Box \phi^2 \nn \\
\label{eq:A_S_int_S_int_result}
 & \qquad \quad + \frac{\hbar^3}{46080 \pi^6} \left( 3 R_{\mu \nu \lambda \rho} R^{\mu \nu \lambda \rho} + 12 R_{\mu \nu} R^{\mu \nu} + 5 R^2 + 58 \Box R \right) \biggr] \vol \,.
\end{align}
This is obviously not of the form \eqref{eq:A_m_exp_S_int_wo_trivial}: There are supplementary terms $R \phi^2$ (note the ``wrong'' sign in the second term on the r.h.s.), $\Box \phi^2$, $R^2$, and $\Box R$. We will see below that these disappear upon performing appropriate redefinitions of time-ordered products. For these, also the following results will be relevant, which can be derived in complete analogy:
\begin{align}
\label{eq:A_S_int_phi_2}
 A( S_\ia \otimes \phi^2(x) ) & = \lambdabar \frac{3 \hbar}{2 \pi^2} \xi(x) \phi^2(x) \,, \\
\label{eq:A_phi_2_phi_2}
 A( \phi^2(x) \otimes \phi^2(x') ) & = \frac{\hbar}{4 \pi^2} \xi(x) \delta(x, x') \,.
\end{align}

Let us now finally turn to the redefinition of time-ordered products. We already mentioned the redefinition \eqref{eq:Z_phi_nabla_nabla_phi} of $\cT( \phi \nabla_\mu \nabla_\nu \phi)$ in order to achieve a conserved stress tensor and the absence of the $\Box R$ term in the free trace anomaly. Now we proceed order by order in the interaction and in $\hbar$ to remove trivial elements from the interacting part of the trace anomaly. We recall from \eqref{eq:A_phi_4} that $A(S_\ia)$ is of first order in $\hbar$ and, by \eqref{eq:Xi_L_0_phi}, cohomologically trivial. According to \eqref{eqs:L_i}, there would be two possibilities to remove this anomaly, namely the Lagrangians $R \phi^2 \vol$ and $\phi \Box \phi \vol$. However, using the latter Langrangian turns out to be inconsistent with field independence, as explained in Appendix~\ref{app:Proofs}. Hence, we perform the $\cO(\hbar)$ redefinition
\beq
\label{eq:Z_1_phi_4}
 Z^{(1)}( \phi^4 ) = \frac{\hbar}{48 \pi^2} R \phi^2 \,.
\eeq
By field independence \eqref{eq:TO_FieldIndependence} holding also for $Z$, we must then also have
\beq
 Z^{(1)}(\phi^2) = \frac{\hbar}{288 \pi^2} R \,.
\eeq
By \eqref{eq:Xi_Z}, the new anomaly fulfills
\beq
 \tilde A( \phi^4 ) = \Xi Z( \phi^4 ) + A( \phi^4 ) + A( Z( \phi^4 ) ) - Z( \tilde A( \phi^4 ) ) \,.
\eeq
At first order in $\hbar$, this reads
\beq
 \tilde A^{(1)}( \phi^4 ) = \Xi Z^{(1)}( \phi^4 ) + A^{(1)}( \phi^4) = 0 \,,
\eeq
so we have indeed removed the anomaly at first order in $\hbar$. At second order in $\hbar$, we then have
\beq
 \tilde A^{(2)}( \phi^4) = \Xi Z^{(2)}( \phi^4 ) + A^{(1)}( Z^{(1)}( \phi^4 ) ) = \Xi Z^{(2)}( \phi^4 ) + \frac{\hbar^2}{2304 \pi^4} R \Box \xi \,,
\eeq
where we used \eqref{eq:A_phi_2}. Hence, using the result for $L_3$ in \eqref{eqs:L_i}, we may remove the anomaly of $\phi^4$ by setting
\beq
\label{eq:Z_2_phi_4}
 Z^{(2)}( \phi^4 ) = \frac{\hbar^2}{27648 \pi^4} R^2 \,.
\eeq
As the \rhs is a c-number, this does not entail any further redefinitions by field independence. The above redefinitions of $\cT( \phi^4 )$ and $\cT( \phi^2 )$ correspond to the redefinitions already found in \cite{PinamontiConformal} to achieve conformally invariant Wick powers without derivatives. It is important to note that we have removed the anomaly completely, not only up to total derivatives, i.e., for any cut-off function $\tilde \chi$, not necessarily constant on the support of $\xi$, we have achieved $\tilde A(S_\ia(\tilde \chi)) = 0$, a condition that we used in our argument that the anomaly can be brought to the form \eqref{eq:A_m_exp_S_int_wo_trivial} up to third order in the interaction. However, this condition does not fix the redefinition \eqref{eq:Z_2_phi_4} completely: One could still add a multiple of $C^2$ to the \rhs of \eqref{eq:Z_2_phi_4} without changing $\tilde A(S_\ia(\tilde \chi)) = 0$. As we will see below, this amounts to the freedom of modifying the anomaly coefficient $c_2$ (the contribution to $c$ at second order in the interaction).

In order to preserve conservation of the stress tensor, the redefinition of $\cT(\phi^2)$ entails that also $\cT( \phi \nabla_\mu \nabla_\nu \phi )$ needs to be redefined. Also the absence of a $\Box R$ term in the trace anomaly of the free theory can then still be achieved (possibly by further redefinitions of $\cT ( \phi \nabla_\mu \nabla_\nu \phi)$). However, these redefinitions do not change the trace anomaly of the free theory, i.e., do not redefine $\cT(\phi ( \Box - \frac{1}{6} R) \phi)$. For later convenience, we summarize the relevant redefinitions performed so far (the first one follows from \eqref{eq:Z_phi_nabla_nabla_phi} and the fact that later redefinitions do not redefine $\cT(\phi ( \Box - \frac{1}{6} R) \phi)$):
\begin{align}
 Z( \phi ( \Box - \tfrac{1}{6} R ) \phi ) & = \frac{\hbar}{720 \pi^2} \left( R^{\alpha \beta \gamma \delta} R_{\alpha \beta \gamma \delta} - R^{\alpha \beta} R_{\alpha \beta} + \frac{5}{4} \Box R \right), \\
 Z( \phi^2 ) & = \frac{\hbar}{288 \pi^2} R \,, \\
\label{eq:Z_phi_4}
 Z( \phi^4 ) & = \frac{\hbar}{48 \pi^2} R \phi^2 + \frac{\hbar^2}{27648 \pi^4} R^2 \,.
\end{align}

We now turn to the anomaly at second order in the interaction. The above redefinitions do not affect the anomaly $A^{(1)}(S_\ia \otimes S_\ia)$ at first order in $\hbar$, i.e., we have
\beq
 \tilde A^{(1)}(S_\ia \otimes S_\ia) = \Xi Z^{(1)}( S_\ia \otimes S_\ia) + A^{(1)}(S_\ia \otimes S_\ia) \,.
\eeq
As the anomaly at first order in $\hbar$ is already of the desired form, we set $Z^{(1)}( S_\ia \otimes S_\ia) = 0$ and obtain
\beq
\label{eq:A_1_S_int_S_int_result}
 \tilde A^{(1)}( S_\ia \otimes S_\ia) = \lambdabar \frac{9 \hbar}{\pi^2} S_\ia(\xi) \,.
\eeq
At second order in $\hbar$, we thus have
\begin{align}
\label{eq:A_2_S_int_S_int}
 \tilde A^{(2)}(S_\ia \otimes S_\ia) & = \Xi Z^{(2)}( S_\ia \otimes S_\ia) + A^{(2)}(S_\ia \otimes S_\ia) + 2 A^{(1)}(S_\ia \otimes Z^{(1)}(S_\ia)) \nn \\
 & \qquad - Z^{(1)}(\tilde A^{(1)}(S_\ia \otimes S_\ia)) \,.
\end{align}
With \eqref{eq:A_S_int_S_int_result} and \eqref{eq:A_S_int_phi_2}, we obtain
\beq
 \tilde A^{(2)}(S_\ia \otimes S_\ia) = \Xi Z^{(2)}( S_\ia \otimes S_\ia) + \lambdabar^2 \hbar^2 \int_M \xi \left( \frac{3}{8 \pi^4} \phi \left( \Box - \frac{1}{6} R \right) \phi + \frac{1}{2 \pi^4} \Box \phi^2 \right) \vol \,,
\eeq
i.e., the last two terms on the \rhs of \eqref{eq:A_2_S_int_S_int} have ``flipped'' the $R \phi^2$ term to the conformally coupled value. The $\Box \phi^2$ term can be removed by setting
\beq
 Z^{(2)}(S_\ia \otimes S_\ia) = \lambdabar^2 \frac{\hbar^2}{12 \pi^4} \int_M R \phi^2 \vol \,,
\eeq
so that we remain with
\beq
\label{eq:A_2_S_int_S_int_result}
 \tilde A^{(2)}(S_\ia \otimes S_\ia) = \lambdabar^2 \frac{3 \hbar^2}{8 \pi^4} \int_M \xi \phi \left( \Box - \frac{1}{6} R \right) \phi \vol \,.
\eeq
Finally, we consider the third order in $\hbar$. We have
\begin{multline}
\label{eq:A_3_S_int_S_int}
 \tilde A^{(3)}(S_\ia \otimes S_\ia) = \Xi Z^{(3)}( S_\ia \otimes S_\ia) + A^{(3)}(S_\ia \otimes S_\ia) + A^{(1)}(Z^{(2)}(S_\ia \otimes S_\ia)) \\
 + A^{(1)}(Z^{(1)}(S_\ia) \otimes Z^{(1)}(S_\ia)) - Z^{(2)}(\tilde A^{(1)}(S_\ia \otimes S_\ia)) - Z^{(1)}(\tilde A^{(2)}(S_\ia \otimes S_\ia)) \,.
\end{multline}
With \eqref{eq:A_S_int_S_int_result}, \eqref{eq:A_phi_2_phi_2}, and the above results, we arrive at
\begin{multline}
 \tilde A^{(3)}(S_\ia \otimes S_\ia) = \Xi Z^{(3)}( S_\ia \otimes S_\ia) \\ + \lambdabar^2 \frac{\hbar^3}{46080 \pi^6} \int_M \xi \left( - 21 R_{\mu \nu \lambda \rho} R^{\mu \nu \lambda \rho} + 36 R_{\mu \nu} R^{\mu \nu} - 5 R^2 + 168 \Box R \right) \vol \,,
\end{multline}
The $\Box R$ term can be removed by the redefinition
\beq
 Z^{(3)}(S_\ia \otimes S_\ia) = \lambdabar^2 \frac{28 \hbar^3}{46080 \pi^6} \int R \phi^2 \vol \,,
\eeq
so that we arrive at
\beq
\label{eq:A_3_S_int_S_int_result}
 \tilde A^{(3)}(S_\ia \otimes S_\ia) = \lambdabar^2 \frac{\hbar^3}{15360 \pi^6} \int_M \xi \left( \cE_4 - 8 C^2 \right) \vol \,.
\eeq

In the notation introduced in \eqref{eq:a_Expansion}, the results \eqref{eq:A_1_S_int_S_int_result}, \eqref{eq:A_2_S_int_S_int_result}, \eqref{eq:A_3_S_int_S_int_result} amount to
\begin{align}
 \beta_2 & =\frac{9 \hbar}{\pi^2} \,, &
 \gamma_2 & = \frac{3 \hbar^2}{8 \pi^4} \,, &
 a_2 & = - \frac{\hbar^3}{15360 \pi^6} \,, &
 c_2 & = - \frac{\hbar^3}{1920 \pi^6} \,.
\end{align}
In particular, we have explicitly verified the relation $a_0 \gamma_2 = - a_2$ derived from the consistency condition. For the effective second order contribution $\tilde c_2 = c_2 + c_0 \gamma_2$ to $c$ (recall \eqref{eq:T_int_T_f_with_gamma}), we obtain
\beq
 \tilde c_2 = - \frac{\hbar^3}{3072 \pi^6} \,.
\eeq
To summarize, we thus have on-shell the trace anomaly
\beq
 \cT^\ia( T(f) ) \simeq \int f \left[ - \frac{\hbar}{5760 \pi^2} \cE_4 + \left( \frac{\hbar}{1920 \pi^2} - \frac{\lambda^2}{4!^2} \frac{\hbar^3}{6144 \pi^6} \right) C^2 + \frac{\lambda^2}{4!^2} \frac{9 \hbar}{2 \pi^2} \phi^4 \right] \vol + \cO(\lambda^3) \,.
\eeq
All these results, including the vanishing effective second order contribution $\tilde a_2 = a_2 + a_0 \gamma_2$, coincide with those obtained in \cite{HathrellScalar}. However, the value for $c_2$ is subject to renormalization ambiguities, namely by adding a term $\hbar^2 c' \, C^2$ to the \rhs of \eqref{eq:Z_2_phi_4} (as discussed below that equation), the fifth term on the \rhs of \eqref{eq:A_3_S_int_S_int} yields a supplementary contribution $- \beta_2 c'$ to $c_2$. In other words, the result for $c_2$ (and thus also $\tilde c_2$) is ambiguous.

\section{Conclusion}

We introduced the notion of Weyl anomaly in quantum field theory on curved spacetimes in the framework of locally covariant field theory. We discussed some of its properties and in particular its relation to the trace anomaly in interacting theories. We studied the case of $\phi^4$ theory both from a cohomological perspective and by explicitly computing the trace anomaly up to second order in the interaction. While to this order our results agree with those of \cite{HathrellScalar}, our finding that one can achieve absence of a $\Box \phi^2$ term at third order in the interaction in a purely perturbative setting is in contradiction to the results of \cite{HathrellScalar} (and those of \cite{IrgesKarageorgos}).

We think that the methods and results presented here should be a fruitful starting point for further investigations. Regarding the general framework, the inclusion of gauge theories seems to be very desirable. For these, gauge fixing breaks the invariance of the free action under local scale transformations. However, this breaking proceeds via terms which are exact \wrt the BRST differential. Hence, we expect that one can generalize the definition of the Weyl anomaly by allowing for supplementary BRST (or even BV) exact terms on the \rhs of \eqref{eq:WeylAnomaly} (see \cite{HollandsYM, FrobBV} for a framework incorporating gauge theories in locally covariant field theory). 
A further important general issue concerns theories involving several linear (basic) fields. In our discussion in Section~\ref{sec:WeylAnomaly}, we treated a term $T(\xi)$ occurring in $A(\et{S_\ia})$ differently from the other terms, both regarding the interpretation of the trace anomaly and the characterization of conformal theories. In the presence of more basic fields, several ``equation of motion terms'' could be present, and not necessarily in a linear combination corresponding to the trace $T$ of the stress tensor. It is then not a priori clear how to deal with these.

It might also be worthwhile to continue the explicit computation of the trace anomaly in the $\phi^4$ model. For example, in \cite{HathrellScalar}, the coefficient of $R \phi^2$ in the trace anomaly is proportional to that of the $\Box \phi^2$ term, underlining the close connection of these terms enforced by consistency conditions. Hence, it seems natural to assume that if the coefficient of $\Box \phi^2$ vanishes at $\cO(\lambda^3)$ (as we can achieve), then also the coefficient of $R \phi^2$ vanishes at $\cO(\lambda^4)$. Also the terms $R^2$ and $\Box R$ in \cite{HathrellScalar} involve the quantity $\eta$ which played an awkward role in the coefficient of $\Box \phi^2$ (\cf the discussion in the Introduction). Hence, it would be reassuring to redo the calculations performed in \cite{HathrellScalar} in the framework and using the methods presented here.

\subsection*{Acknowledgements}

J.~Z.~would like to thank Stefan Hollands for useful discussions. We thank Leonidas Karageorgos for pointing out \cite{IrgesKarageorgos} and discussions on that work, and the anonymous reviewers for their careful reading of the manuscript. This work is supported by the Deutsche Forschungsgemeinschaft (DFG, German Research Foundation) --- project no. 396692871 within the Emmy Noether grant CA1850/1-1.

\appendix

\section{Proofs of some statements}
\label{app:Proofs}

We sketch the proof of the basic relation \eqref{eq:A_f}, in particular that $A_f( F_1 \otimes \dots F_k)$ is local both in the $F_j$ and in $f$. We proceed as in the proof of Lemma~8 in \cite{HollandsYM}, namely by induction in the number $k$ of factors of local functionals. For $k=1$, we have to consider
\beq
 A_f(F) = \cT^{-1} ( \delta^\weyl_f \cT( F ) - \cT( \delta^\weyl_f F ) ) \,.
\eeq
A basic fact is that application of $\delta^\weyl_f$ on a local functional yields a local functional with support contained in the intersection of $\supp F$ and $\supp f$. 
Now for any local functional $F$, also $\cT(F)$ and $\cT^{-1}(F)$ are local functionals which agree with $F$ at $\cO(\hbar^0)$. 
Hence, $A_f(F)$ as defined above is a local functional, linear in $f$ (as $\delta^\weyl_f$ is linear in $f$), supported within $\supp f \cap \supp F$, and of $\cO(\hbar)$. 
This provides the induction start. Now assume that we have proven the desired statement
\beq
 \delta^\weyl_f \cT( F_1 \otimes \dots F_l ) = \sum_{j = 1}^l \cT( F_1 \otimes \dots \delta^\weyl_f F_j \otimes \dots F_l ) + \sum_{j = 1}^l \sum_{I_j} \left( \ibar \right)^{1-j} \cT( A_f( F_{I_j} ) \otimes F_{I^{\mathrm{c}}_j} ) \,,
\eeq
for all $l < k$. 
Here the sum over $I_j$ refers to all subsets of $\{ 1, \dots, l \}$ of length $j$, and $F_I$ stands for $F_{i_1} \otimes \dots F_{i_j}$ with $i_k$ the elements of $I$. 
Furthermore, $I_j^{\mathrm{c}}$ stands for the complementary subset of $\{ 1, \dots l \}$. In order to prove the induction step, we define $A_f(F_1 \otimes \dots F_k)$ as the ``missing term'' in an analogous expansion of $\delta^\weyl_f \cT(F_1 \otimes \dots F_k)$, i.e.,
\begin{multline}
\label{eq:A_f_F_1_F_k}
 A_f(F_1 \otimes \dots F_k) \defeq \left( \ibar \right)^{k-1} \cT^{-1} \left[ \delta^\weyl_f \cT(F_1 \otimes \dots F_k) - \sum_{j = 1}^k \cT( F_1 \otimes \dots \delta^\weyl_f F_j \otimes \dots F_k ) \right. \\
 \left.- \sum_{j = 1}^{k-1} \sum_{I_j} \left( \ibar \right)^{1-j} \cT( A_f( F_{I_j} ) \otimes F_{I^{\mathrm{c}}_j} ) \right].
\end{multline}
To show that the expression in the large brackets is a local functional, assume that not all supports of the $F_j$ overlap. 
By a partition of unity argument based on \cite[Lemma~4.1]{BrunettiFredenhagenScalingDegree}, one can then split the local functionals $F_j$ as finite sums $F_j = F_{j,1} + F_{j,2} + \dots$ of local functionals, such that any set $\{ F_{1, i_1}, \dots, F_{k, i_k} \}$ can be split into two non-empty subsets so that the support of any element of the first subset does not intersect the causal past of the support of any element of the second subset. The contribution of any such set $\{ F_{1, i_1}, \dots, F_{k, i_k} \}$ to $A_f(F_1 \otimes \dots F_k)$ according to \eqref{eq:A_f_F_1_F_k} can then be shown to vanish using causal factorization and the inductive assumption. Hence, $A_f(F_1 \otimes \dots F_k)$ must be supported on $\cap_j \supp F_j$, i.e., it must be a local functional.
The same equation can be shown to also hold for connected time-ordered products as defined in \cite[Proof of Lemma~8]{HollandsYM} or \cite{FrobBV}, and by considering that equation, one finds that $A_f( F_1 \otimes \dots F_k )$ is of $\cO(\hbar)$, in complete analogy to \cite[Lemma~8]{HollandsYM} or \cite[Theorem~3]{FrobBV}.
It remains to show that the expression in large brackets is also local in $f$, i.e., vanishes if $\supp f$ does not intersect $\cap_j \supp F_j$. But that follows from the inductive assumption, the locality of $\delta^\weyl_f F_j$, and the local covariance of time-ordered products (for any $x$ there is an arbitrarily small neighborhood $U$ such that $\cT( \Phi_1(x_1) \otimes \dots \Phi_k(x_k))$ for $x_j \in U$ is independent of the geometric data outside of $U$ for arbitrary fields $\Phi_j$).

To prove the behaviour \eqref{eq:Xi_sigma} of the world function $\sigma$ under conformal transformations, we evaluate the definition \eqref{eq:sigma} of the world function in terms of normal coordinates around $x'$, such that the geodesic from $x$ to $x'$ is given by $z^\mu(\tau) = x^\mu - \tau \chi^\mu$ with the normal vector $\chi^\mu = x^\mu - (x')^\mu$. 
As geodesics extremize the energy functional \eqref{eq:sigma}, the variation of the geodesic does (at first order) not contribute to the variation of the world function, only the variation of the metric contributes.
Furthermore, we have
\beq
 \chi^\mu g_{\mu \nu}(z(\tau)) = \chi^\mu g_{\mu \nu}(x') = \chi^\mu \eta_{\mu \nu} \,,
\eeq
so that (the last equality is to be understood in the sense of an asymptotic expansion)
\begin{align}
 \Xi \sigma(x, x') & = \int_0^1 \xi(z(\tau)) g_{\mu \nu}(z(\tau)) \frac{\ud z^\mu}{\ud \tau} \frac{\ud z^\nu}{\ud \tau} \ud \tau \nn \\
 & = \chi^\mu \chi^\nu \eta_{\mu \nu} \int_0^1 \xi( x - \tau \chi ) \ud \tau \nn \\
 & = 2 \sigma(x, x') \sum_{k = 0}^\infty \chi^{\mu_1} \dots \chi^{\mu_k} \del_{\mu_1} \dots \del_{\mu_k} \xi(x') \int_0^1 \frac{(-1)^k \tau^k}{k!} \ud \tau + \cO( (\sigma^\mu)^\infty ),
\end{align}
which gives the first line of \eqref{eq:Xi_sigma} upon realizing that, in Riemannian normal coordinates around $x$, $\del_{\mu_1} \dots \del_{\mu_k} \xi(x)$ coincides with $\nabla_{(\mu_1} \dots \nabla_{\mu_k)} \xi(x)$ \cite[Eqn.~(2.5)]{FrobOPE}.

We now show that the Lagrangian $\phi \Box \phi \vol$ cannot be used for a redefinition of $\cT(\phi^4)$ in order to achieve $A(\phi^4) = 0$, i.e., we cannot set
\beq
\label{eq:Z_phi_4_impossible}
 Z( \phi^4(\tilde \chi) ) = c \int \tilde \chi \phi \Box \phi \vol \,.
\eeq
where $\tilde \chi$ is an arbitrary test function and $\phi^4(\tilde \chi)$ stands for integration of $\phi^4$ with this test function. Namely by functionally differentiating \wrt $\phi$ in the direction $\vp$ and using field independence \eqref{eq:TO_FieldIndependence}, we obtain
\beq
 Z( \phi^3( \vp \tilde \chi )) = \frac{c}{4} \int \left( \Box ( \tilde \chi \vp ) + \tilde \chi \Box \vp \right) \phi \vol \,.
\eeq 
While the \lhs depends on $\tilde \chi$ and $\vp$ only through their product $\vp \tilde \chi$, the \rhs cannot be written as depending only on $\vp \tilde \chi$ (and its derivatives). Hence, the redefinition \eqref{eq:Z_phi_4_impossible} is not possible.

\bibliography{../../mybib}{}

\providecommand{\href}[2]{#2}\begingroup\raggedright\begin{thebibliography}{10}

\bibitem{CapperDuff1974}
D.M.~Capper and M.J.~Duff, \emph{{Trace anomalies in dimensional
  regularization}}, \href{https://doi.org/10.1007/BF02748300}{\emph{Nuovo Cim.
  A} {\bfseries 23} (1974) 173}.

\bibitem{Christensen76}
S.~Christensen, \emph{{Vacuum Expectation Value of the Stress Tensor in an
  Arbitrary Curved Background: The Covariant Point Separation Method}},
  \href{https://doi.org/10.1103/PhysRevD.14.2490}{\emph{Phys.\ Rev.} {\bfseries
  D14} (1976) 2490}.

\bibitem{Brown1977}
L.S.~Brown, \emph{{Stress Tensor Trace Anomaly in a Gravitational Metric:
  Scalar Fields}}, \href{https://doi.org/10.1103/PhysRevD.15.1469}{\emph{Phys.
  Rev. D} {\bfseries 15} (1977) 1469}.

\bibitem{ChristensenFulling77}
S.M.~Christensen and S.A.~Fulling, \emph{{Trace Anomalies and the Hawking
  Effect}}, \href{https://doi.org/10.1103/PhysRevD.15.2088}{\emph{Phys. Rev.}
  {\bfseries D15} (1977) 2088}.

\bibitem{DowkerCritchley}
J.S.~Dowker and R.~Critchley, \emph{{The Stress Tensor Conformal Anomaly for
  Scalar and Spinor Fields}},
  \href{https://doi.org/10.1103/PhysRevD.16.3390}{\emph{Phys. Rev. D}
  {\bfseries 16} (1977) 3390}.

\bibitem{Duff1977}
M.J.~Duff, \emph{{Observations on Conformal Anomalies}},
  \href{https://doi.org/10.1016/0550-3213(77)90410-2}{\emph{Nucl. Phys. B}
  {\bfseries 125} (1977) 334}.

\bibitem{WaldTraceAnomaly}
R.M.~Wald, \emph{{Trace Anomaly of a Conformally Invariant Quantum Field in
  Curved Space-Time}},
  \href{https://doi.org/10.1103/PhysRevD.17.1477}{\emph{Phys.\ Rev.} {\bfseries
  D17} (1978) 1477}.

\bibitem{DuffReview1993}
M.J.~Duff, \emph{{Twenty years of the Weyl anomaly}},
  \href{https://doi.org/10.1088/0264-9381/11/6/004}{\emph{Class. Quant. Grav.}
  {\bfseries 11} (1994) 1387}
  [\href{https://arxiv.org/abs/hep-th/9308075}{{\ttfamily hep-th/9308075}}].

\bibitem{1702.04247}
F.~Bastianelli, O.~Corradini and E.~Vassura, \emph{{Quantum mechanical path
  integrals in curved spaces and the type-A trace anomaly}},
  \href{https://doi.org/10.1007/JHEP04(2017)050}{\emph{JHEP} {\bfseries 04}
  (2017) 050} [\href{https://arxiv.org/abs/1702.04247}{{\ttfamily
  1702.04247}}].

\bibitem{CasarinGodazgarNicolai}
L.~Casarin, H.~Godazgar and H.~Nicolai, \emph{{Conformal Anomaly for
  Non-Conformal Scalar Fields}},
  \href{https://doi.org/10.1016/j.physletb.2018.10.034}{\emph{Phys. Lett. B}
  {\bfseries 787} (2018) 94}
  [\href{https://arxiv.org/abs/1809.06681}{{\ttfamily 1809.06681}}].

\bibitem{BastianelliChiese}
F.~Bastianelli and L.~Chiese, \emph{{Chiral fermions, dimensional
  regularization, and the trace anomaly}},
  \href{https://doi.org/10.1016/j.nuclphysb.2022.115914}{\emph{Nucl. Phys. B}
  {\bfseries 983} (2022) 115914}
  [\href{https://arxiv.org/abs/2203.11668}{{\ttfamily 2203.11668}}].

\bibitem{FilocheLarueQuevillonVuong}
B.~Filoche, R.~Larue, J.~Quevillon and P.N.H.~Vuong, \emph{{Anomalies from an
  effective field theory perspective}},
  \href{https://doi.org/10.1103/PhysRevD.107.025017}{\emph{Phys. Rev. D}
  {\bfseries 107} (2023) 025017}
  [\href{https://arxiv.org/abs/2205.02248}{{\ttfamily 2205.02248}}].

\bibitem{CorianoLionettiMaglio}
C.~Corian\`o, S.~Lionetti and M.M.~Maglio, \emph{{CFT correlators and
  CP-violating trace anomalies}},
  \href{https://doi.org/10.1140/epjc/s10052-023-11984-z}{\emph{Eur. Phys. J. C}
  {\bfseries 83} (2023) 839}
  [\href{https://arxiv.org/abs/2307.03038}{{\ttfamily 2307.03038}}].

\bibitem{LarueQuevillonZwicky}
R.~Larue, J.~Quevillon and R.~Zwicky, \emph{{Trace anomaly of weyl fermions via
  the path integral}},
  \href{https://doi.org/10.1007/JHEP12(2023)064}{\emph{JHEP} {\bfseries 12}
  (2023) 064} [\href{https://arxiv.org/abs/2309.08670}{{\ttfamily
  2309.08670}}].

\bibitem{FerreroEtAlTraceAnomaly}
R.~Ferrero, S.A.~Franchino-Vi\~nas, M.B.~Fr\"ob and W.C.C.~Lima,
  \emph{{Universal Definition of the Nonconformal Trace Anomaly}},
  \href{https://doi.org/10.1103/PhysRevLett.132.071601}{\emph{Phys. Rev. Lett.}
  {\bfseries 132} (2024) 071601}
  [\href{https://arxiv.org/abs/2312.07666}{{\ttfamily 2312.07666}}].

\bibitem{2312.13222}
R.~Larue, J.~Quevillon and R.~Zwicky, \emph{{Gravity-gauge anomaly constraints
  on the energy-momentum tensor}},
  \href{https://doi.org/10.1007/JHEP05(2024)307}{\emph{JHEP} {\bfseries 05}
  (2024) 307} [\href{https://arxiv.org/abs/2312.13222}{{\ttfamily
  2312.13222}}].

\bibitem{2406.12464}
T.~Bertolini and L.~Casarin, \emph{{Conformal anomalies and renormalized stress
  tensor correlators for nonconformal theories}},
  \href{https://doi.org/10.1103/PhysRevD.110.045007}{\emph{Phys. Rev. D}
  {\bfseries 110} (2024) 045007}
  [\href{https://arxiv.org/abs/2406.12464}{{\ttfamily 2406.12464}}].

\bibitem{2407.00415}
G.~Paci and O.~Zanusso, \emph{{Weyl cohomology and the conformal anomaly in the
  presence of torsion}},
  \href{https://doi.org/10.1016/j.aop.2024.169877}{\emph{Annals Phys.}
  {\bfseries 473} (2025) 169877}
  [\href{https://arxiv.org/abs/2407.00415}{{\ttfamily 2407.00415}}].

\bibitem{2411.03842}
G.~Paci and O.~Zanusso, \emph{{Ambient space and integration of the trace
  anomaly}}, \href{https://doi.org/10.1007/JHEP03(2025)111}{\emph{JHEP}
  {\bfseries 03} (2025) 111}
  [\href{https://arxiv.org/abs/2411.03842}{{\ttfamily 2411.03842}}].

\bibitem{Nakayama2012}
Y.~Nakayama, \emph{{CP-violating CFT and trace anomaly}},
  \href{https://doi.org/10.1016/j.nuclphysb.2012.02.006}{\emph{Nucl. Phys. B}
  {\bfseries 859} (2012) 288}
  [\href{https://arxiv.org/abs/1201.3428}{{\ttfamily 1201.3428}}].

\bibitem{Bonora2014}
L.~Bonora, S.~Giaccari and B.~Lima~de Souza, \emph{{Trace anomalies in chiral
  theories revisited}},
  \href{https://doi.org/10.1007/JHEP07(2014)117}{\emph{JHEP} {\bfseries 07}
  (2014) 117} [\href{https://arxiv.org/abs/1403.2606}{{\ttfamily 1403.2606}}].

\bibitem{Bastianelli2016}
F.~Bastianelli and R.~Martelli, \emph{{On the trace anomaly of a Weyl
  fermion}}, \href{https://doi.org/10.1007/JHEP11(2016)178}{\emph{JHEP}
  {\bfseries 11} (2016) 178}
  [\href{https://arxiv.org/abs/1610.02304}{{\ttfamily 1610.02304}}].

\bibitem{FrobZahn2019}
M.B.~Fr\"ob and J.~Zahn, \emph{{Trace anomaly for chiral fermions via Hadamard
  subtraction}}, \href{https://doi.org/10.1007/JHEP10(2019)223}{\emph{JHEP}
  {\bfseries 10} (2019) 223}
  [\href{https://arxiv.org/abs/1904.10982}{{\ttfamily 1904.10982}}].

\bibitem{AbdallahEtAl2021}
S.~Abdallah, S.A.~Franchino-Vi\~nas and M.B.~Fr\"ob, \emph{{Trace anomaly for
  Weyl fermions using the Breitenlohner-Maison scheme for $\gamma_*$}},
  \href{https://doi.org/10.1007/JHEP03(2021)271}{\emph{JHEP} {\bfseries 03}
  (2021) 271} [\href{https://arxiv.org/abs/2101.11382}{{\ttfamily
  2101.11382}}].

\bibitem{HorowitzWald1978}
G.T.~Horowitz and R.M.~Wald, \emph{{Dynamics of Einstein's Equation Modified by
  a Higher Order Derivative Term}},
  \href{https://doi.org/10.1103/PhysRevD.17.414}{\emph{Phys. Rev. D} {\bfseries
  17} (1978) 414}.

\bibitem{BonoraEtAl83}
L.~Bonora, P.~Cotta-Ramusino and C.~Reina, \emph{{Conformal Anomaly and
  Cohomology}}, \href{https://doi.org/10.1016/0370-2693(83)90169-7}{\emph{Phys.
  Lett. B} {\bfseries 126} (1983) 305}.

\bibitem{BrownCollins}
L.S.~Brown and J.C.~Collins, \emph{{Dimensional Renormalization of Scalar Field
  Theory in Curved Space-time}},
  \href{https://doi.org/10.1016/0003-4916(80)90232-8}{\emph{Annals Phys.}
  {\bfseries 130} (1980) 215}.

\bibitem{HathrellScalar}
S.J.~Hathrell, \emph{{Trace Anomalies and $\lambda \phi^4$ Theory in Curved
  Space}}, \href{https://doi.org/10.1016/0003-4916(82)90008-2}{\emph{Annals
  Phys.} {\bfseries 139} (1982) 136}.

\bibitem{HathrellQED}
S.J.~Hathrell, \emph{{Trace Anomalies and {QED} in Curved Space}},
  \href{https://doi.org/10.1016/0003-4916(82)90227-5}{\emph{Annals Phys.}
  {\bfseries 142} (1982) 34}.

\bibitem{Freeman1983}
M.D.~Freeman, \emph{{The Renormalization of Nonabelian Gauge Theories in Curved
  Space-time}},
  \href{https://doi.org/10.1016/0003-4916(84)90022-8}{\emph{Annals Phys.}
  {\bfseries 153} (1984) 339}.

\bibitem{Osborn1989}
H.~Osborn, \emph{{Derivation of a Four-dimensional $c$ Theorem}},
  \href{https://doi.org/10.1016/0370-2693(89)90729-6}{\emph{Phys. Lett. B}
  {\bfseries 222} (1989) 97}.

\bibitem{JackOsborn}
I.~Jack and H.~Osborn, \emph{{Analogs for the $c$ Theorem for Four-dimensional
  Renormalizable Field Theories}},
  \href{https://doi.org/10.1016/0550-3213(90)90584-Z}{\emph{Nucl. Phys. B}
  {\bfseries 343} (1990) 647}.

\bibitem{Osborn1991}
H.~Osborn, \emph{{Weyl consistency conditions and a local renormalization group
  equation for general renormalizable field theories}},
  \href{https://doi.org/10.1016/0550-3213(91)80030-P}{\emph{Nucl. Phys. B}
  {\bfseries 363} (1991) 486}.

\bibitem{Cardy1988}
J.L.~Cardy, \emph{{Is There a c Theorem in Four-Dimensions?}},
  \href{https://doi.org/10.1016/0370-2693(88)90054-8}{\emph{Phys. Lett. B}
  {\bfseries 215} (1988) 749}.

\bibitem{KomargodskiSchwimmer}
Z.~Komargodski and A.~Schwimmer, \emph{{On Renormalization Group Flows in Four
  Dimensions}}, \href{https://doi.org/10.1007/JHEP12(2011)099}{\emph{JHEP}
  {\bfseries 12} (2011) 099} [\href{https://arxiv.org/abs/1107.3987}{{\ttfamily
  1107.3987}}].

\bibitem{Fortin:2012hn}
J.-F.~Fortin, B.~Grinstein and A.~Stergiou, \emph{{Limit Cycles and Conformal
  Invariance}}, \href{https://doi.org/10.1007/JHEP01(2013)184}{\emph{JHEP}
  {\bfseries 01} (2013) 184} [\href{https://arxiv.org/abs/1208.3674}{{\ttfamily
  1208.3674}}].

\bibitem{HollandsWaldWick}
S.~Hollands and R.M.~Wald, \emph{{Local Wick polynomials and time ordered
  products of quantum fields in curved space-time}},
  \href{https://doi.org/10.1007/s002200100540}{\emph{Commun.\ Math.\ Phys.}
  {\bfseries 223} (2001) 289}
  [\href{https://arxiv.org/abs/gr-qc/0103074}{{\ttfamily gr-qc/0103074}}].

\bibitem{HollandsWaldTO}
S.~Hollands and R.M.~Wald, \emph{{Existence of local covariant time ordered
  products of quantum fields in curved space-time}},
  \href{https://doi.org/10.1007/s00220-002-0719-y}{\emph{Commun.\ Math.\ Phys.}
  {\bfseries 231} (2002) 309}
  [\href{https://arxiv.org/abs/gr-qc/0111108}{{\ttfamily gr-qc/0111108}}].

\bibitem{BrunettiFredenhagenVerch}
R.~Brunetti, K.~Fredenhagen and R.~Verch, \emph{{The Generally covariant
  locality principle: A New paradigm for local quantum field theory}},
  {\emph{Commun.\ Math.\ Phys.} {\bfseries 237} (2003) 31}
  [\href{https://arxiv.org/abs/math-ph/0112041}{{\ttfamily math-ph/0112041}}].

\bibitem{HollandsWaldReview}
S.~Hollands and R.M.~Wald, \emph{{Quantum fields in curved spacetime}},
  \href{https://doi.org/10.1016/j.physrep.2015.02.001}{\emph{Phys.\ Rept.}
  {\bfseries 574} (2015) 1} [\href{https://arxiv.org/abs/1401.2026}{{\ttfamily
  1401.2026}}].

\bibitem{HaagBook}
R.~Haag, \emph{Local quantum physics}, Springer-Verlag, Berlin, second~ed.
  (1996),
  \href{https://doi.org/10.1007/978-3-642-61458-3}{10.1007/978-3-642-61458-3}.

\bibitem{HollandsWaldRG}
S.~{Hollands} and R.M.~{Wald}, \emph{{On the Renormalization Group in Curved
  Spacetime}}, \href{https://doi.org/10.1007/s00220-003-0837-1}{\emph{Commun.\
  Math.\ Phys.} {\bfseries 237} (2003) 123}
  [\href{https://arxiv.org/abs/gr-qc/0209029}{{\ttfamily gr-qc/0209029}}].

\bibitem{CosteriDappiaggiGoi}
B.~Costeri, C.~Dappiaggi and M.~Goi, \emph{{Conservation Law and Trace Anomaly
  for the Stress Energy Tensor of a Self-Interacting Scalar Field}},
  \href{https://doi.org/10.1007/s00023-025-01580-0}{\emph{Ann. H. Poincar{\'e}}
  (2025) } [\href{https://arxiv.org/abs/2411.07109}{{\ttfamily 2411.07109}}].

\bibitem{HollandsYM}
S.~Hollands, \emph{{Renormalized Quantum Yang-Mills Fields in Curved
  Spacetime}}, \href{https://doi.org/10.1142/S0129055X08003420}{\emph{Rev.\
  Math.\ Phys.} {\bfseries 20} (2008) 1033}
  [\href{https://arxiv.org/abs/0705.3340}{{\ttfamily 0705.3340}}].

\bibitem{IrgesKarageorgos}
N.~Irges and L.~Karageorgos, \emph{{Energy-Momentum tensor correlators in
  \ensuremath{\phi^4} theory I: The spin-zero sector}},
  \href{https://doi.org/10.1016/j.nuclphysb.2024.116782}{\emph{Nucl. Phys. B}
  {\bfseries 1010} (2025) 116782}
  [\href{https://arxiv.org/abs/2410.16040}{{\ttfamily 2410.16040}}].

\bibitem{HollandsWaldStress}
S.~Hollands and R.M.~Wald, \emph{{Conservation of the stress tensor in
  interacting quantum field theory in curved spacetimes}},
  \href{https://doi.org/10.1142/S0129055X05002340}{\emph{Rev.\ Math.\ Phys.}
  {\bfseries 17} (2005) 227}
  [\href{https://arxiv.org/abs/gr-qc/0404074}{{\ttfamily gr-qc/0404074}}].

\bibitem{WaldGR}
R.M.~Wald, \emph{General Relativity}, University of Chicago Press (1984).

\bibitem{LocCovDirac}
J.~Zahn, \emph{{The renormalized locally covariant Dirac field}},
  \href{https://doi.org/10.1142/S0129055X13300124}{\emph{Rev.\ Math.\ Phys.}
  {\bfseries 26} (2014) 1330012}
  [\href{https://arxiv.org/abs/1210.4031}{{\ttfamily 1210.4031}}].

\bibitem{DuetschFredenhagenLoopExpansion}
M.~D{\"u}tsch and K.~Fredenhagen, \emph{{Algebraic quantum field theory,
  perturbation theory, and the loop expansion}},
  \href{https://doi.org/10.1007/PL00005563}{\emph{Commun.\ Math.\ Phys.}
  {\bfseries 219} (2001) 5}
  [\href{https://arxiv.org/abs/hep-th/0001129}{{\ttfamily hep-th/0001129}}].

\bibitem{Radzikowski}
M.~Radzikowski, \emph{{Micro-local approach to the Hadamard condition in
  quantum field theory on curved space-time}},
  \href{https://doi.org/10.1007/BF02100096}{\emph{Commun.\ Math.\ Phys.}
  {\bfseries 179} (1996) 529}.

\bibitem{PinamontiConformal}
N.~Pinamonti, \emph{{Conformal generally covariant quantum field theory: The
  Scalar field and its Wick products}},
  \href{https://doi.org/10.1007/s00220-009-0780-x}{\emph{Commun. Math. Phys.}
  {\bfseries 288} (2009) 1117}
  [\href{https://arxiv.org/abs/0806.0803}{{\ttfamily 0806.0803}}].

\bibitem{BreDue}
F.~Brennecke and M.~D{\"u}tsch, \emph{{Removal of violations of the Master Ward
  Identity in perturbative QFT}}, {\emph{Rev.\ Math.\ Phys.} {\bfseries 20}
  (2008) 119} [\href{https://arxiv.org/abs/0705.3160}{{\ttfamily 0705.3160}}].

\bibitem{BackgroundIndependence}
J.~Zahn, \emph{{Locally covariant charged fields and background independence}},
  \href{https://doi.org/10.1142/S0129055X15500178}{\emph{Rev.\ Math.\ Phys.}
  {\bfseries 27} (2015) 1550017}
  [\href{https://arxiv.org/abs/1311.7661}{{\ttfamily 1311.7661}}].

\bibitem{BDF09}
R.~Brunetti, M.~D{\"u}tsch and K.~Fredenhagen, \emph{{Perturbative Algebraic
  Quantum Field Theory and the Renormalization Groups}}, {\emph{Adv.\ Theor.\
  Math.\ Phys.} {\bfseries 13} (2009) 1541}
  [\href{https://arxiv.org/abs/0901.2038}{{\ttfamily 0901.2038}}].

\bibitem{EpsteinGlaser}
H.~Epstein and V.~Glaser, \emph{{The role of locality in perturbation theory}},
  {\emph{Ann.\ Inst.\ H.\ Poincar{\'e}} {\bfseries 19} (1973) 211}.

\bibitem{BrunettiFredenhagenScalingDegree}
R.~Brunetti and K.~Fredenhagen, \emph{{Microlocal analysis and interacting
  quantum field theories: Renormalization on physical backgrounds}},
  \href{https://doi.org/10.1007/s002200050004}{\emph{Commun.\ Math.\ Phys.}
  {\bfseries 208} (2000) 623}
  [\href{https://arxiv.org/abs/math-ph/9903028}{{\ttfamily math-ph/9903028}}].

\bibitem{Lowenstein71}
J.H.~Lowenstein, \emph{{Differential vertex operations in Lagrangian field
  theory}}, \href{https://doi.org/10.1007/BF01907030}{\emph{Commun.\ Math.\
  Phys.} {\bfseries 24} (1971) 1}.

\bibitem{Lam72}
Y.-M.P.~Lam, \emph{{Perturbation Lagrangian theory for scalar fields:
  Ward-Takahashi identity and current algebra}},
  \href{https://doi.org/10.1103/PhysRevD.6.2145}{\emph{Phys.\ Rev.} {\bfseries
  D6} (1972) 2145}.

\bibitem{BreitenlohnerMaison}
P.~Breitenlohner and D.~Maison, \emph{{Dimensional Renormalization and the
  Action Principle}}, \href{https://doi.org/10.1007/BF01609069}{\emph{Commun.
  Math. Phys.} {\bfseries 52} (1977) 11}.

\bibitem{KhavkineMoretti}
I.~Khavkine and V.~Moretti, \emph{{Analytic Dependence is an Unnecessary
  Requirement in Renormalization of Locally Covariant QFT}},
  \href{https://doi.org/10.1007/s00220-016-2618-7}{\emph{Commun. Math. Phys.}
  {\bfseries 344} (2016) 581}
  [\href{https://arxiv.org/abs/1411.1302}{{\ttfamily 1411.1302}}].

\bibitem{MoroPhD}
A.~Moro, \emph{{Functional formalism for algebraic classical and quantum field
  theories}}, Ph.D. thesis, Trento U., 2023.
\newblock \href{https://arxiv.org/abs/2308.04856}{{\ttfamily 2308.04856}}.

\bibitem{HofmannPhD}
A.~Hofmann, \emph{{Microlocal methods in quantum field theory}}, Ph.D. thesis,
  G{\"o}ttingen U., 2023.
\newblock 10.53846/goediss-10857.

\bibitem{AlvarezGaumeWitten}
L.~Alvarez-Gaume and E.~Witten, \emph{{Gravitational Anomalies}},
  \href{https://doi.org/10.1016/0550-3213(84)90066-X}{\emph{Nucl.\ Phys.}
  {\bfseries B234} (1984) 269}.

\bibitem{PopineauStora}
G.~Popineau and R.~Stora, \emph{{A pedagogical remark on the main theorem of
  perturbative renormalization theory}},
  \href{https://doi.org/10.1016/j.nuclphysb.2016.04.046}{\emph{Nucl. Phys. B}
  {\bfseries 912} (2016) 70}.

\bibitem{DuetschFredenhagenAWI}
M.~D{\"u}tsch and K.~Fredenhagen, \emph{{Causal perturbation theory in terms of
  retarded products, and a proof of the action Ward identity}},
  \href{https://doi.org/10.1142/S0129055X04002266}{\emph{Rev.\ Math.\ Phys.}
  {\bfseries 16} (2004) 1291}
  [\href{https://arxiv.org/abs/hep-th/0403213}{{\ttfamily hep-th/0403213}}].

\bibitem{AdlerBardeenThm}
J.~Zahn, \emph{{Background Independence and the Adler--Bardeen Theorem}},
  \href{https://doi.org/10.1007/s00023-023-01368-0}{\emph{Annales Henri
  Poincar{\'e}} {\bfseries 25} (2024) 2641}
  [\href{https://arxiv.org/abs/2209.02393}{{\ttfamily 2209.02393}}].

\bibitem{Habil}
J.~Zahn, \emph{{Anomalies in quantized gauge field theories}}, habilitation
  thesis, submitted to Leipzig University, 2025.

\bibitem{IlinSlavnov}
V.~{Il'in} and D.A.~Slavnov, \emph{{Algebras of observables in the S-matrix
  approach}}, \href{https://doi.org/10.1007/BF01035870}{\emph{Theor. Math.
  Phys.} {\bfseries 36} (1978) 578}.

\bibitem{FrobBV}
M.B.~Fr\"ob, \emph{{Anomalies in Time-Ordered Products and Applications to the
  BV--BRST Formulation of Quantum Gauge Theories}},
  \href{https://doi.org/10.1007/s00220-019-03558-6}{\emph{Commun. Math. Phys.}
  {\bfseries 372} (2019) 281}
  [\href{https://arxiv.org/abs/1803.10235}{{\ttfamily 1803.10235}}].

\bibitem{BgInd}
M.~Taslimi~Tehrani and J.~Zahn, \emph{{Background independence in gauge
  theories}}, \href{https://doi.org/10.1007/s00023-020-00887-4}{\emph{Annales
  Henri Poincar{\'e}} {\bfseries 21} (2020) 1135}
  [\href{https://arxiv.org/abs/1804.07640}{{\ttfamily 1804.07640}}].

\bibitem{RejznerFredenhagenQuantization}
K.~Fredenhagen and K.~Rejzner, \emph{{Batalin-Vilkovisky formalism in
  perturbative algebraic quantum field theory}},
  \href{https://doi.org/10.1007/s00220-012-1601-1}{\emph{Commun.\ Math.\ Phys.}
  {\bfseries 317} (2013) 697}
  [\href{https://arxiv.org/abs/1110.5232}{{\ttfamily 1110.5232}}].

\bibitem{deMedeirosHollands}
P.~de~Medeiros and S.~Hollands, \emph{{Superconformal quantum field theory in
  curved spacetime}},
  \href{https://doi.org/10.1088/0264-9381/30/17/175015}{\emph{Class. Quant.
  Grav.} {\bfseries 30} (2013) 175015}
  [\href{https://arxiv.org/abs/1305.0499}{{\ttfamily 1305.0499}}].

\bibitem{BabujianKarowski}
H.~Babujian and M.~Karowski, \emph{{Sine-Gordon breather form-factors and
  quantum field equations}},
  \href{https://doi.org/10.1088/0305-4470/35/43/308}{\emph{J. Phys. A}
  {\bfseries 35} (2002) 9081}
  [\href{https://arxiv.org/abs/hep-th/0204097}{{\ttfamily hep-th/0204097}}].

\bibitem{AlbertiSchlesierZahn}
M.A.A.~Martin, R.~Schlesier and J.~Zahn, \emph{{Semiclassical energy density of
  kinks and solitons}},
  \href{https://doi.org/10.1103/PhysRevD.107.065002}{\emph{Phys. Rev. D}
  {\bfseries 107} (2023) 065002}
  [\href{https://arxiv.org/abs/2204.08785}{{\ttfamily 2204.08785}}].

\bibitem{FrobCadamuroSineGordon}
M.B.~Fr\"ob and D.~Cadamuro, \emph{{Local operators in the Sine-Gordon model:
  $\partial_\mu \phi \, \partial_\nu \phi$ and the stress tensor}},
  \href{https://doi.org/10.1007/s00023-025-01565-z}{\emph{Ann. H. Poincar{\'e}}
  (2025) } [\href{https://arxiv.org/abs/2205.09223}{{\ttfamily 2205.09223}}].

\bibitem{MorettiStressEnergy}
V.~Moretti, \emph{{Comments on the stress energy tensor operator in curved
  space-time}}, \href{https://doi.org/10.1007/s00220-002-0702-7}{\emph{Commun.\
  Math.\ Phys.} {\bfseries 232} (2003) 189}
  [\href{https://arxiv.org/abs/gr-qc/0109048}{{\ttfamily gr-qc/0109048}}].

\bibitem{PoissonPoundVega}
E.~Poisson, A.~Pound and I.~Vega, \emph{{The Motion of point particles in
  curved spacetime}}, \href{https://doi.org/10.12942/lrr-2011-7}{\emph{Living
  Rev. Rel.} {\bfseries 14} (2011) 7}
  [\href{https://arxiv.org/abs/1102.0529}{{\ttfamily 1102.0529}}].

\bibitem{DecaniniFolacci08}
Y.~Decanini and A.~Folacci, \emph{{Hadamard renormalization of the
  stress-energy tensor for a quantized scalar field in a general spacetime of
  arbitrary dimension}},
  \href{https://doi.org/10.1103/PhysRevD.78.044025}{\emph{Phys.\ Rev.}
  {\bfseries D78} (2008) 044025}
  [\href{https://arxiv.org/abs/gr-qc/0512118}{{\ttfamily gr-qc/0512118}}].

\bibitem{Steinmann}
O.~Steinmann, \emph{Perturbation expansions in axiomatic field theory}, vol.~11
  of \emph{Lecture Notes in Physics}, Springer-Verlag, Berlin-New York (1971).

\bibitem{BahnsWrochna}
D.~Bahns and M.~Wrochna, \emph{{On-shell extension of distributions}},
  \href{https://doi.org/10.1007/s00023-013-0288-y}{\emph{Annales Henri
  Poincare} {\bfseries 15} (2014) 2045}
  [\href{https://arxiv.org/abs/1210.5448}{{\ttfamily 1210.5448}}].

\bibitem{DangExtension}
N.V.~Dang, \emph{{The Extension of Distributions on Manifolds, a Microlocal
  Approach}}, \href{https://doi.org/10.1007/s00023-015-0419-8}{\emph{Annales
  Henri Poincare} {\bfseries 17} (2016) 819}
  [\href{https://arxiv.org/abs/1412.2808}{{\ttfamily 1412.2808}}].

\bibitem{FreedmanJohnsonLatorre}
D.Z.~Freedman, K.~Johnson and J.I.~Latorre, \emph{{Differential regularization
  and renormalization: A New method of calculation in quantum field theory}},
  \href{https://doi.org/10.1016/0550-3213(92)90240-C}{\emph{Nucl. Phys. B}
  {\bfseries 371} (1992) 353}.

\bibitem{xact}
J.M.~Mart{\'\i}n-Garc{\'\i}a et~al., ``{xAct: Efficient tensor computer algebra
  for the Wolfram Language}.'' \href{http://www.xact.es}{http://www.xact.es},
  2024.

\bibitem{FrobOPE}
M.B.~Fr{\"o}b, \emph{{Recursive construction of the operator product expansion
  in curved space}}, \href{https://doi.org/10.1007/JHEP02(2021)195}{\emph{JHEP}
  {\bfseries 02} (2021) 195}
  [\href{https://arxiv.org/abs/2007.15668}{{\ttfamily 2007.15668}}].

\end{thebibliography}\endgroup
\bibliographystyle{../../JHEP}

\end{document}